\documentclass[twocolumn,twoside,slac_two]{revtex4}
\usepackage{graphicx}
\usepackage{fancyhdr}
\input{boonesym}
\def\likee{\ensuremath{\mathcal L}_e \xspace}
\def\likemu{\ensuremath{\mathcal L}_\mu \xspace}
\def\likepi{\ensuremath{\mathcal L}_\pi \xspace}
\def\logemu{\ensuremath{\log(\likee/\likemu)\xspace}}
\def\logepi{\ensuremath{\log(\likee/\likepi)\xspace}}
\def\mgg{\ensuremath{m_{\gamma\gamma}}}
\def\enuqe{\ensuremath{E_\nu^{QE}}}

\begin{document}

\title{The Search for $\num\to\nue$ Oscillations at MiniBooNE}

%

\author{H. A. Tanaka, for the MiniBooNE collaboration}
\affiliation{Department of Physics, Joseph Henry Laboratories, 
Princeton University, Princeton, New Jersey, 08544 United States of America}

\begin{abstract}
MiniBooNE (Mini Booster Neutrino Experiment) searches for the $\num\to\nue$
oscillations with $\Delta m^2 \sim 1 \evsq$ indicated by the LSND experiment. The LSND evidence, when taken with the solar and atmospheric neutrino oscillations,  suggests new physics beyond
the Standard Model. However, this evidence
has not been confirmed by other experiments. MiniBooNE 
has completed its first $\num\to\nue$ oscillation search using 
a sample of $\sim\!\!1\gev$ neutrino events obtained with $5.58\times 10^{20}$ protons delivered to the Booster Neutrino Beamline.  The analysis finds no significant excess of $\nue$ events in the analysis region of $475-3000\mev$.
\end{abstract}

\maketitle

\thispagestyle{fancy}


\section{Introduction}
The long-standing deficits in observed solar electron neutrinos ($\nue$) 
\cite{homestake,sage,gallex,superksolar,snocc,snonc,snosalt} and atmospheric muon
neutrinos ($\num$) \cite{kamioka_atm1,kamioka_atm2,superk_atm,soudan_atm,macro_atm}
now have a firm explanation in terms of the flavor oscillations of the three neutrinos in the Standard Model. This interpretation has been confirmed  with experiments using terrestial neutrino sources\cite{kamland1, kamland2,k2k,minos} and point to oscillations with
mass-squared differences of $\Delta m_{12}^2 \!\sim\! 8\times 10^{-5} \evsq$ and $\Delta m _{23}^2\sim3\times 10^{-3}\evsq$ for the solar and atmospheric oscillations, respectively. However, the indications for $\numb\to\nueb$ oscillations reported by the LSND collaboration with $\Delta m^2 \!\sim\! \mathcal{O}(1\ev^2)$\cite{lsnd}  cannot be accommodated within this picture. Taken with other constraints, such as the
number of light neutrinos with standard weak couplings obtained from  measurements of the $Z$ width\cite{l3_nnu,aleph_nnu}, this result  would require a dramatic departure from the Standard Model in the form of ``sterile'' neutrinos
without standard weak couplings and/or exotic forms of symmetry violation\cite{sterile}. The LSND results have not been confirmed, though other experiments searching for neutrino oscillations
with similar values of $\Delta m^2$ have not had sufficient sensitivity to rule it out completely\cite{karmen,bugey}.

The Mini Booster Neutrino Experiment (MiniBooNE) searches for the neutrino oscillations 
indicated by the LSND result using an $\mathcal{O}(1\gev)$ $\num$ beam produced by 1.6-$\mus$-long
pulses of $8\gev$ protons from the Fermilab Booster synchrotron\cite{beamtdr}. Typically,
$\sim 4\times 10^{12}$ protons are delivered in such a pulse at a rate of $\sim 4$ Hz. The protons
in each spill are measured by two toroids before they impinge on a 71-cm-long beryllium target, where the $p$-Be interactions produce secondary mesons (dominated by pions, with a small contribution from kaons), which in turn decay in a 50-meter-long decay region following the target to produce the neutrino beam. The beryllium target is embedded within an electromagnet (``horn'') pulsed synchronously with the beam with a 174 kA current to produce a toroidal magnetic field that focuses positive(negative) particles, resulting in a $\nu(\nub)$-enhanced beam. In its first results, MiniBooNE has performed a search for $\num\to\nue$ oscillations  using $5.58\times 10^{20}$ protons-on-target collected in neutrino mode\cite{mbprl}. The resulting neutrino beam is $>99\%$ pure in $\num$ and peaks at $\sim 700\mev$, with a small component of $\nue$ coming from three-body decays of muons and kaons. 

The MiniBooNE detector \cite{boonetdr} is situated 541 meters downstream of the target, near the axis defined
by the beam. This distance, along with the energy of the $\num$ beam,  matches the distance/energy ratio of the LSND experiment ($\sim\!0.8 \mbox{m}/\mev$), with the result that the oscillation probability is similar in the two experiments. The detector is a 12.2 meter diameter
sphere filled with 800 tons of undoped Marcol 7 mineral oil with an index of refraction of $\sim\!\!1.47$. The sphere is divided into two concentric, optically decoupled regions at 575 cm radius, resulting in  ``main'' and ``veto'' regions. The main region is instrumented on its
outer surface with 1280 inward-facing 8'' photomultiplier tubes (PMTs) covering $10\%$ of the
surface. The veto region is instrumented with 240 8'' PMTs. Neutrino interactions within the main region are identified via the Cherenkov radiation produced by the charged particles emerging from the
interaction and detected on the PMT array.  Cosmic muons entering the main region can be identified and rejected by the light produced and detected in the veto region. The PMT activity in a 19.2
$\mus$ window around each batch of protons delivered from the Booster is recorded, where the expected arrival time of the neutrinos occurs 4.6 to 6.2 $\mus$ after the start of the window. Other activity in the detector, in the form of random triggers and calibration data, is recorded to study
the detector response and non-beam backgrounds.

\begin{figure*}[t]
\centering
\includegraphics[width=82.5 mm]{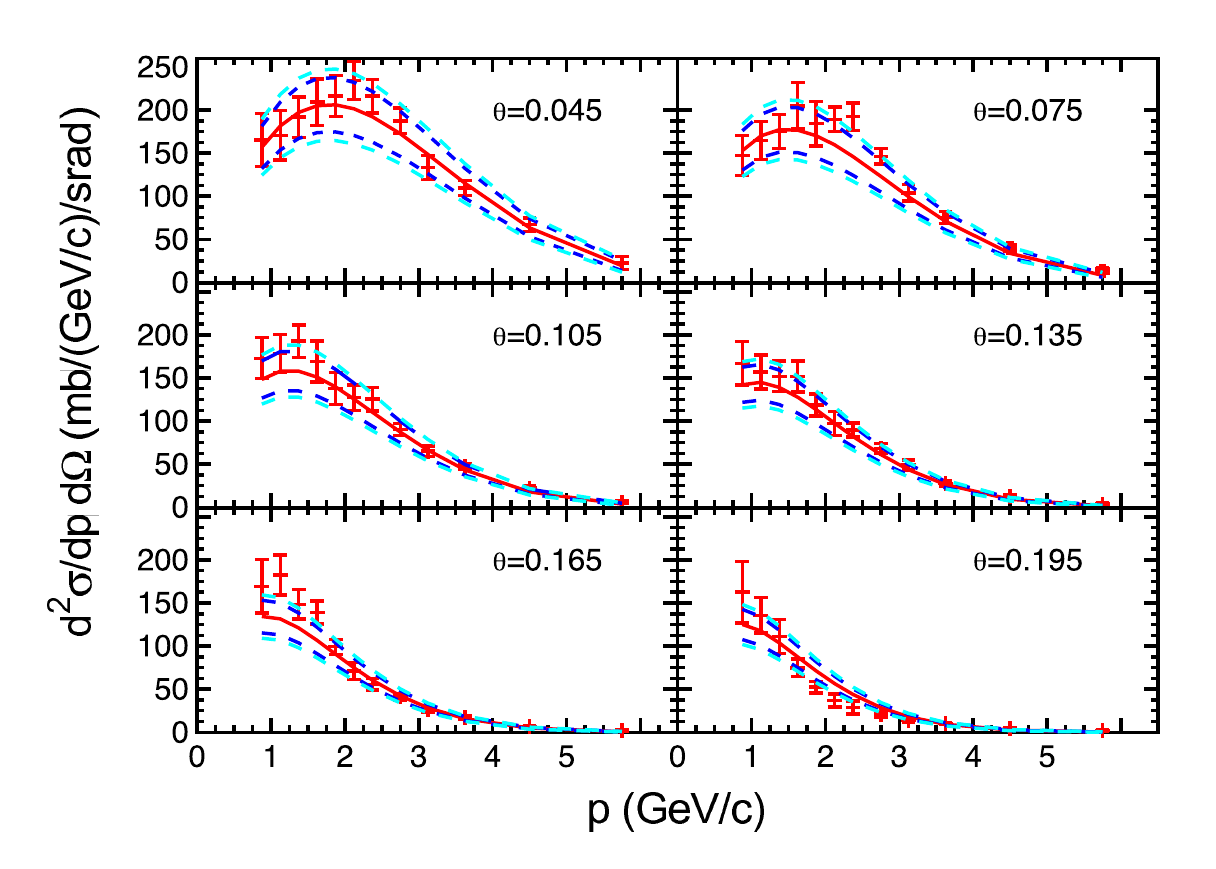}
\includegraphics[width=82.5 mm]{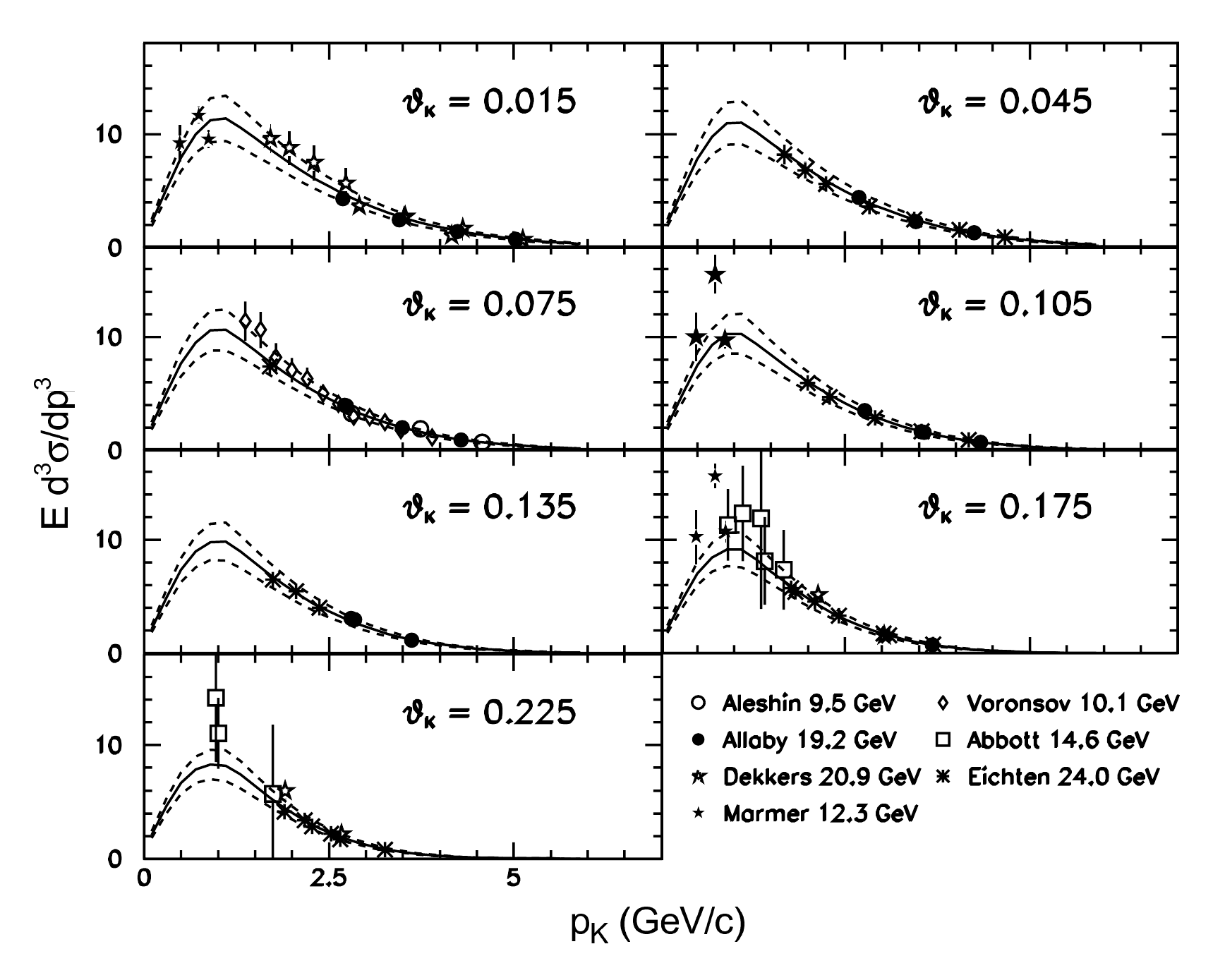}
\caption{Absolute particle production rates in $8\gev$ $p$-Be interactions for $\pip$ (left)
 and $\Kp$ (right). The $\pip$ measurements show the
double differential cross sections measured by the HARP experiment as a function
of $\pip$ momentum in bins of the polar angle of the $\pip$ relative to the incoming proton. 
The $\Kp$ measurements show invariant cross sections measured with proton energies  of 9.5-24 
$\gev$ and  extrapolated to 8 $\gev$ using the Feynman scaling hypothesis. 
The solid red(black) lines for $\pip(\Kp)$ show the parametrization obtained by fitting the
data, while the dashed lines show the variation in the parametrized cross sections
when the parameters are varied according to their uncertainties.} \label{fig:pionkaon}
\end{figure*}
  
\section{The Booster Neutrino Beam}
The predicted neutrino flux at the detector is obtained from
a GEANT4-based Monte Carlo simulation\cite{geant4}. The simulation includes
a detailed description of the beamline geometry, the spatial and kinematic properties of 
the primary proton beam incident of the target, and pion and kaon production models
tuned to the available data. The primary protons and secondary particles are tracked 
through the beamline geometry, accounting for the magnetic field due to the horn, as well as electromagnetic and hadronic interactions in the material (which can produce further particles), until a particle decays to produce a neutrino. In the case where a decay chain has multiple opportunities
to produce neutrinos ({\em e.g.} $\pip\to\mup+\num$, $\mup\to \ep+\numb+\nue$), all decays producing neutrinos are recorded.

\begin{figure*}[t]
\centering
\includegraphics[width=100.0 mm]{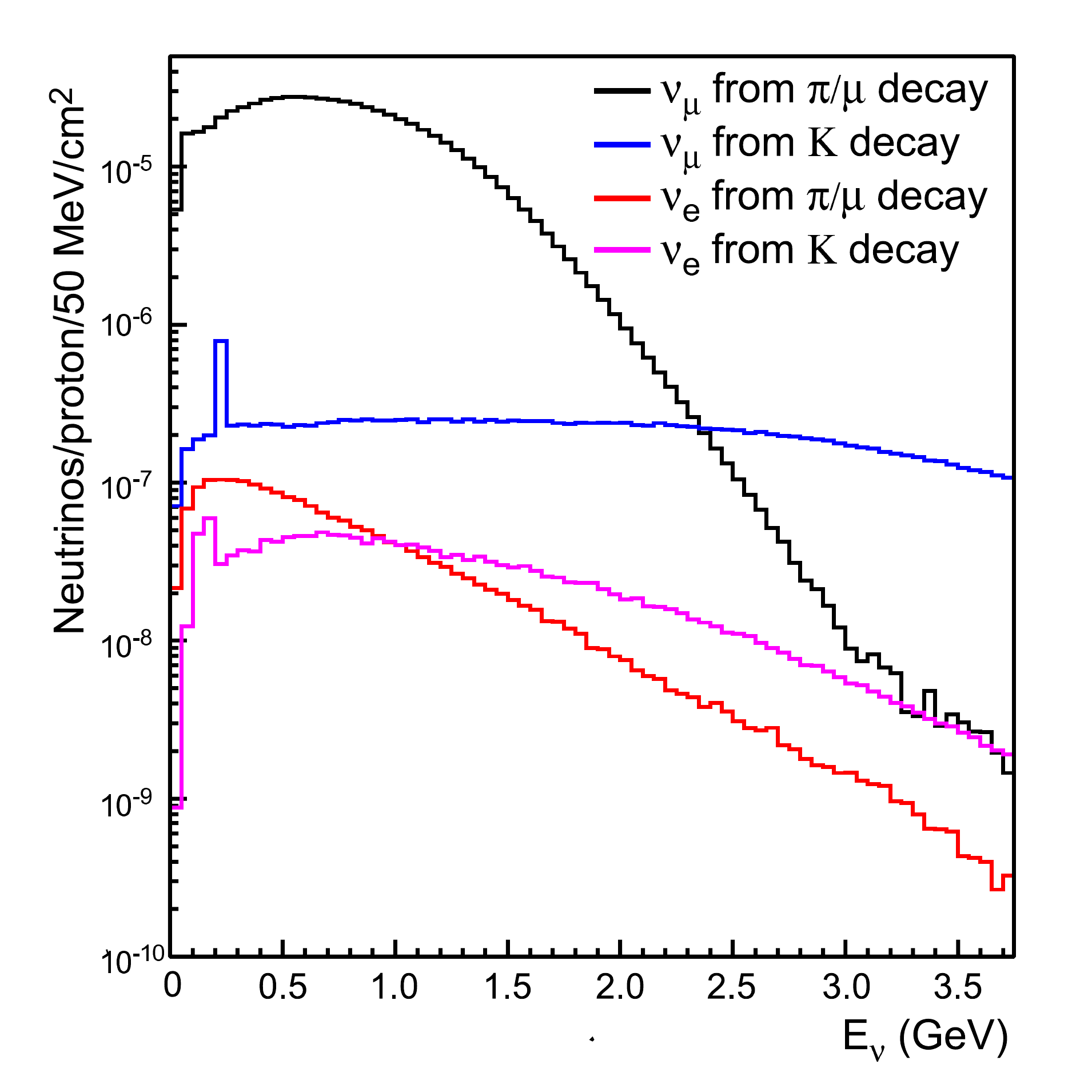}
\caption{The neutrino flux at the MiniBooNE detector predicted by a GEANT4-based Monte Carlo simulation of the Booster Neutrino Beamline.} \label{fig:flux}
\end{figure*}

The primary source of the neutrino flux is the decay of $\pip$ from $p$-Be interactions. The HARP experiment at CERN has measured the momentum and angular
distribution of $\pipm$ production for $8\gev$ protons on a thin beryllium target\cite{harp}. Similar measurements
at 6 $\gevc$ and $12.3\gevc$ proton momentum are available from the BNL E910 experiment. These measurements are summarized as absolute double differential cross sections in pion momentum
and angle (relative to the incident proton) and parametrized using the Sanford-Wang function\cite{sw}. The parameters of the function are obtained via a fit to the three datasets.
The results are shown on the left in Figure \ref{fig:pionkaon}, where the parametrization (in red) is overlayed on the $\pip$  double differential cross sections from the HARP experiment. The
cyan curves show the variations in the function when the underlying parameters are varied according to the uncertainties returned by the fit. The
function is used directly in the simulation to determine the multiplicity and kinematics of 
$\pip$ mesons emerging from $p$-Be interactions. A similar procedure is used for the $\pim$
production based on the HARP and E910 data, while an E910 analysis of $\KS$ production is used
to obtain a Sanford-Wang parametrization of $\Kz$ production.

The $\Kp$ production in $p$-Be interactions has been measured by a number
of experiments. However, no measurements at  $8\gev$ proton energy
exist. As a result, measurements at energies between
9.5 and 24 $\gev$\cite{kaons} are used, with the Lorentz-invariant cross sections extrapolated to $8\gev$ using an assumption of scaling in the Feynman $x$ variable\cite{fs}. The scaled invariant cross sections are parametrized as a function of $x$, with the results shown on the right in  Figure \ref{fig:pionkaon}. 
As with the $\pip$ Sanford-Wang parametrization, this Feynman scaling-based function is used
directly in the simuation to determine the multiplicity and kinematics of the outgoing $\Kp$ in 
$p$-Be interactions.

Cross sections for three categories of hadron-nuclear interactions (elastic, quasi-elastic, and inelastic scattering) have been adjusted to match existing measurements for proton, neutron and charged pion on beryllium and aluminum\cite{bobchenko,gachurin,allardyce,ashery}. In some cases, theoretical guidance \cite{glauber} is used to infer cross sections where direct measurements do not exist.  In all other hadron-nucleus interactions, the GEANT4 default cross sections are used. 

The predicted neutrino flux at the MiniBooNE detector is shown in Figure \ref{fig:flux}.
The dominant part of the $\num$ flux, particularly at $<2\gev$, is due to pion decays, while the kaon contribution becomes larger for energies $>2.5\gev$.  For the $\nue$ flux,
muon decay is the largest component, while kaon decays become dominant at $>1\gev$. A number of systematic uncertainties in the flux prediction are considered. The largest
uncertainty arises from the modeling of the secondary particle production, while uncertainties
in the hadronic cross sections also significant. The $\num$ flux at high energy
is particularly sensitive to the details of the modeling of the horn magnetic field as well as
the hadronic cross sections. While systematic uncertainties in  the predicted  neutrino flux, together with neutrino interaction cross sections,  lead to large uncertainties in  the predicted rate of both background and signal processes, {\em in situ} measurements of the event rates of a number of neutrino-induced processes (discussed in Section \ref{sec:eventrate})  reduce the impact of  these uncertainties.

\section{The MiniBooNE detector}
The primary means of (charged) particle detection in MiniBooNE is via
the Cherenkov radiation produced by such particles as they traverse the mineral oil. A full understanding of the detector response to such particles also requires understanding the scintillation processes
which constitute another source of light production, as 
well as the various processes which the optical photons from both sources undergo prior to their
detection by the PMTs. A detailed understanding of the properties of the charge and
time response of the PMTs to these photons is also needed. 
The measurements and constraints described in the following sections are incorporated
into a GEANT3-based Monte Carlo simulation \cite{geant3} of the detector.

\begin{figure*}[t]
\centering
\includegraphics[width=80mm]{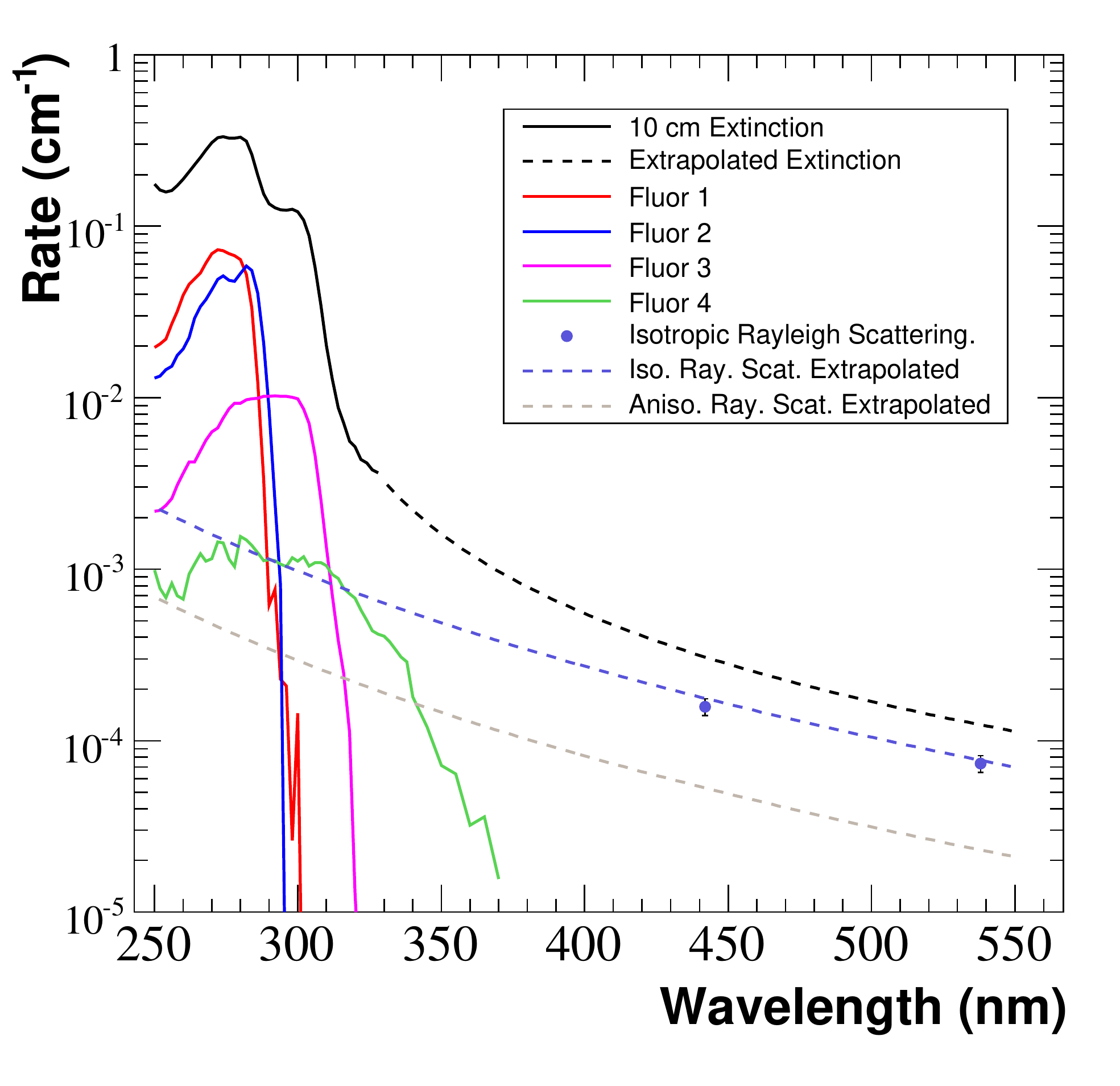}
\includegraphics[width=80mm]{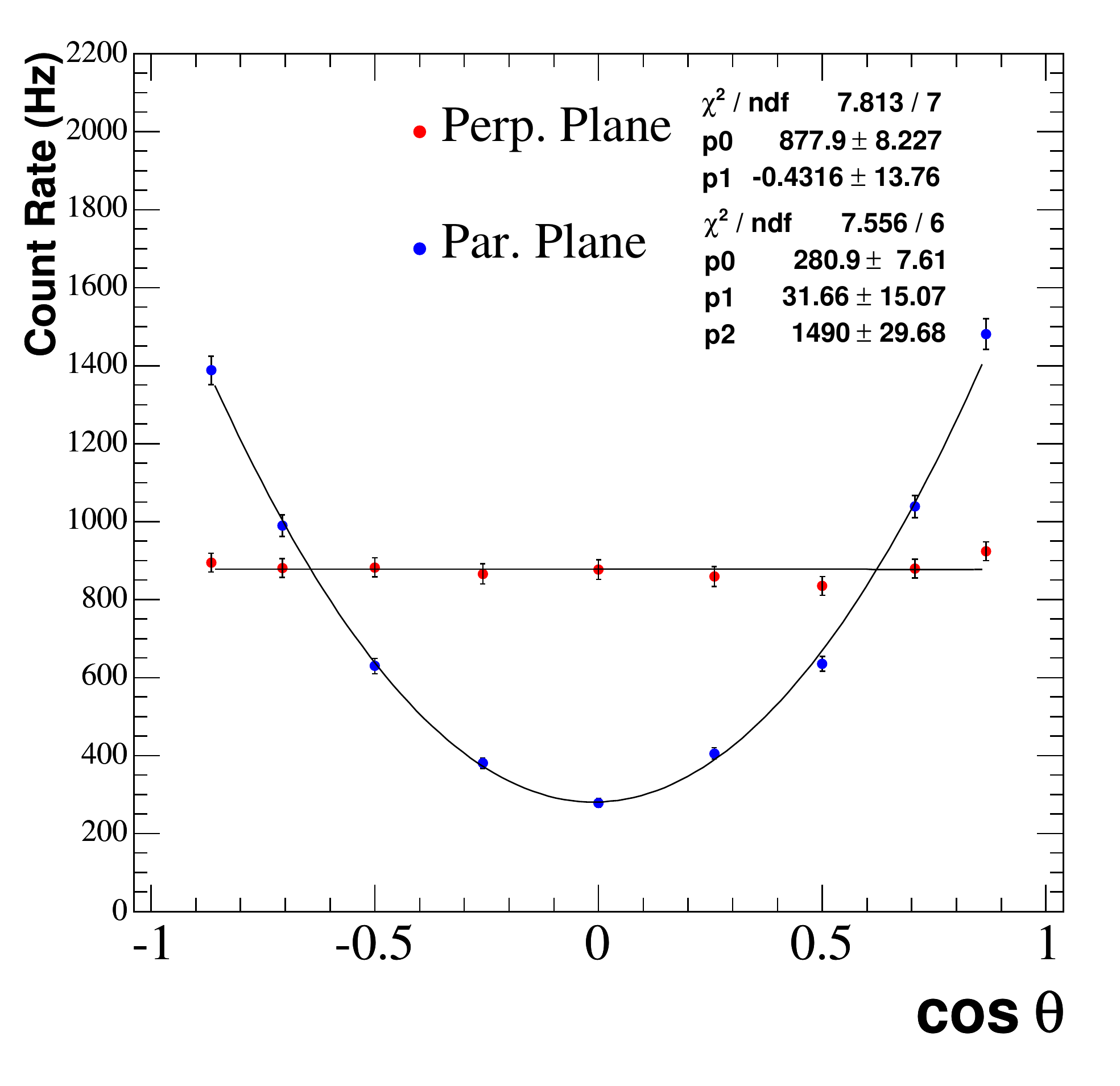}
\caption{Left: Rates of various optical processes in Marcol 7 from
external measurements as a function of wavelength. The solid black line is the
overall extinction rates obtained from spectrophotometer measurements through a 10 cm cell. 
The dashed black line is the extrapolated extinction rate based on {\em in situ} data. The 
curves labled ``Fluor 1-4'' are the measured excitation rates for the four identified fluorescence processes. The lavender points represent the measured rate of Rayleigh scattering at 442 and 538 nm,
with the dashed lavender and gray lines representing the theoretically extrapolated rates. Right:
The measured angular distribution of Rayleigh scattering at 442 nm in the plane perpendicular (red)
and parallel (blue) to the incident polarization. The latter allows the extraction of the contributions
from isotropic and anisotropic fluctuations.}  \label{fig:exsitu}
\end{figure*}

\subsection{External Measurements on Marcol 7}
While the Marcol 7 mineral oil used in MiniBooNE is remarkably transparent 
(extinction lengths of over 30 meters at 400 nm, near the peak of the photocathode sensitivity of
 the PMTs), it also exhibits a rich array of optical phenomena. In addition
 to the Cherenkov and scintillation processes which generate light, these include
 fluorescence processes with different excitation/emission spectra and lifetimes,
 Rayleigh and Raman scattering, and absorption. The left plot in Figure \ref{fig:exsitu} summarizes
the rate of the various processes as a function of wavelength. The cumulative extinction
rate (the sum of all optical processes) is shown as the black line.
In the near ultraviolet region ($<320\nm$), a number of fluorescence processes dominate, 
leading to a large increase in the extinction rate in this region. In the visible region ($>320\nm$),
the dominant processes are Rayleigh scattering and absorption.

The optical processes are identified and studied in {\em ex situ} studies
using small samples of Marcol 7 with $1-10 \cm$ path lengths\cite{brown}. The index of refraction,
which summarizes the dielectric properties of the mineral oil relevant for Cherenkov
radiation and Rayleigh scattering, has been measured using a sodium lamp ($589.3\nm$)
and the observed dispersion used to parametrize the wavelength dependence. The extrapolated
wavelength dependence is verified with direct measurements at wavelengths between $400-700\nm$. The wavelength dependence of the overall extinction rate is measured  by analyzing the transmission rate through Marcol 7 samples using a spectrophotometer. 

The angular dependence of scattering in Marcol 7 is determined by measuring the intensity of light scattered from 442 and 532 nm lasers as a function of angle using a PMT as shown on the right in Figure \ref{fig:exsitu}. Together with its dependence on the polarization of the incident and scattered light, the scattering is found to be consistent with Rayleigh scattering from both isotropic and anisotropic
thermal density fluctuations. The absolute rate of the scattering was determined by calibrating the measured PMT
rates using suspensions of 50 nm polysterene spheres with known number density using the
scattering cross section calculated from Mie theory\cite{bohren}. 

The excitation and emission spectra of the fluorescence properties are determined using steady-state spectrophotometer
measurements at $250-600\nm$ wavelength on 1 cm samples of Marcol 7. A singular-value decomposition analysis identified four major components of fluorescence. Time-resolved measurements, which use a pulsed dye laser emitting ultraviolet light at wavelengths between 285 and 310 nm with $7\ps$ width, decomposed the observed time distribution of emitted light into components with different emission lifetimes as a function of emission wavelength. The result is the emission
spectrum for each identified lifetime component. By matching these emission spectra
to those determined from the steady-state measurements, a complete model of
each fluorescence process (excitation spectrum, and spectrum and lifetime of emission) is
obtained. Finally, the steady state measurements also identified a Raman scattering process.

\begin{figure*}[t]
\centering
\includegraphics[width=80 mm]{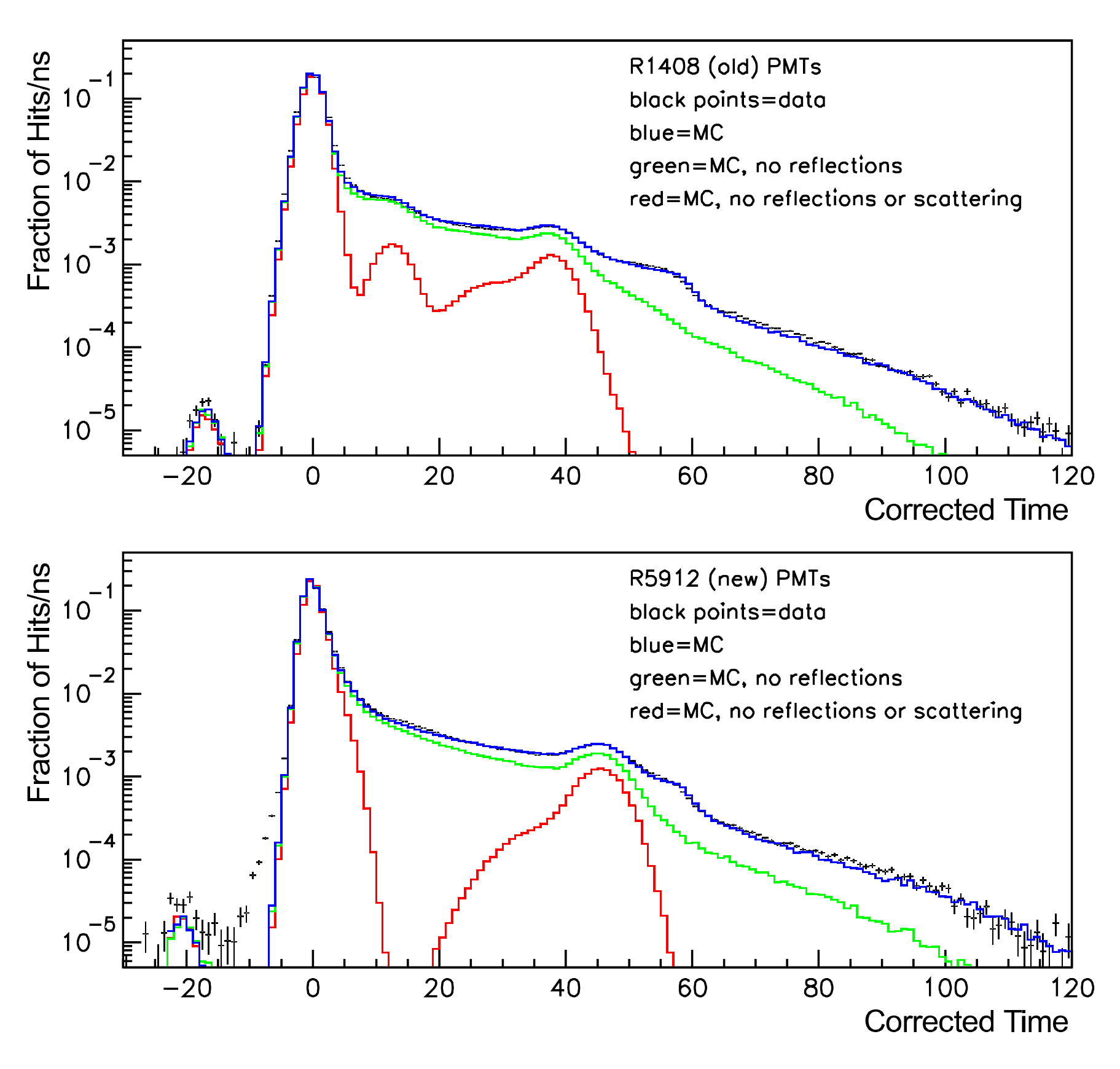}
\includegraphics[width=80mm]{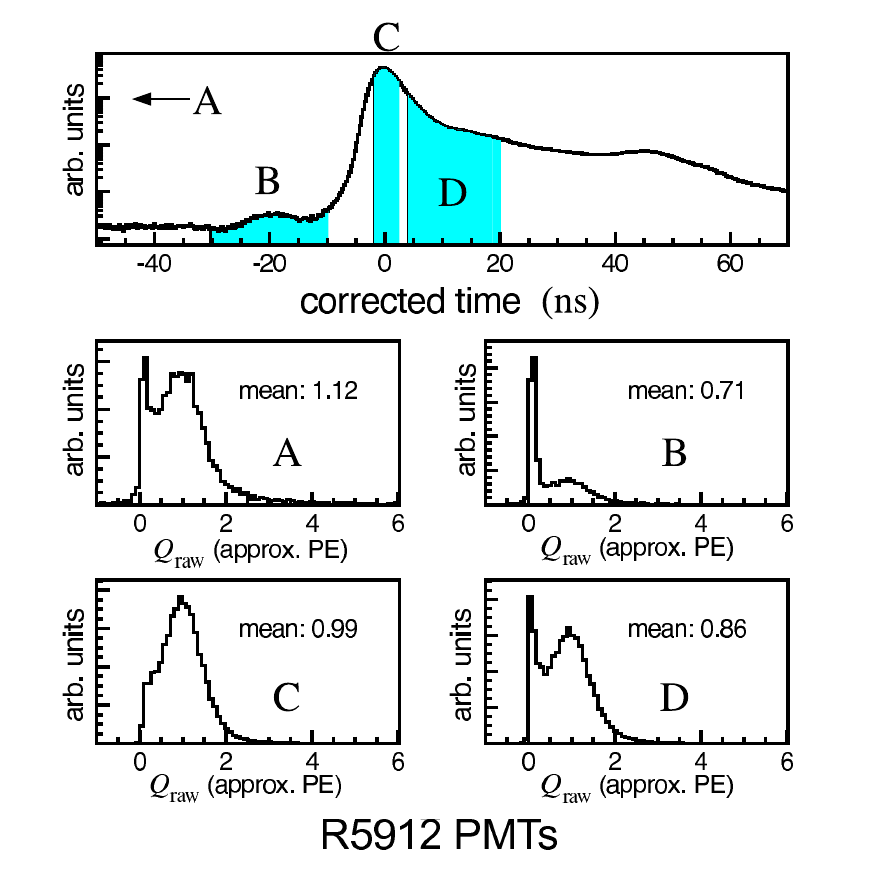}

\caption{Left: Reconstructed photon arrival times for R1408 PMTs (top) and R5912 PMTs (bottom) for light flashed from the center laser flask.  The black histogram is the distribution from data , while the red represents the Monte Carlo simulation with reflections and scattering suppressed to isolate the time structure from the PMT response. A flat background level has been subtracted from the data distribution based on the observed rate before the expected arrival time of the photons. The green and blue histograms the effects of turning on
the scattering and reflections, respectively so that the latter includes all known optical processes. 
Right: Correlations between the reconstructed time and charge observed in center laser flask
data observed in R5912 PMTs. The four bottom plots show the reconstructed charged distribution observed in the four regions of reconstructed photon arrival times indicated in the top plot.} \label{fig:laser}
\end{figure*}

\subsection{Photomultiplier Response}
The 1520 8'' PMTs in  MiniBooNE are of two types: 322 model R5912 PMTs
and 1198 model R1408 PMTs, both from Hamamatsu. All of the  R5912 PMTs are located
in the main array, while the veto array is composed entirely of R1408 PMTs.
A detailed understanding of the PMT response to optical photons is needed
to represent the response of the detector to neutrino events accurately. The time response
of the PMT plays a critical role in the tuning of the optical properties of Marcol 7 using
{\em in situ} data described in Section \ref{sec:omtune}. 

The response of the PMTs to optical photons is characterized by both external
and {\em in situ} calibration data. The wavelength-dependent efficiency of the photocathode 
has been measured by the manufacturer. The time response for a small sample of  PMTs, particularly
the late-pulsing behavior, along with the variation of the efficiency with incident angle, has been characterized in external measurements using a pulsed LED \cite{pmt}. 

Within the MiniBooNE detector, the response of the PMTs is studied using
a laser calibration system. The system consists of four glass flasks situated within the
main region (including one at the center) containing colloidal silica (Ludox\textregistered\cite{ludox}) to disperse light entering the flasks via optical fibers coupled to pulsed
diode lasers at 397 and 438 nm wavelength outside the detector. The primary purpose of the system is to provide gain and time-offset calibrations via the charges measured in low-intensity pulses and the reconstructed times from the center flask. The low intensity minimizes photon pileup so that the
pulses recorded by the PMTs correspond to single photoelectrons. 
Due to the controlled nature of both the geometry and timing of the photon emission,
the center laser flask system proved to be a valuable tool for analyzing the details of the PMT
response. Among these features include the tube-to-tube variations in the delay of the late pulsing and
the correlations between the recorded charge and reconstructed time relative to the expected arrival
time.  

The left plot on Figure \ref{fig:laser} shows the arrival time of hits recorded in laser events relative to their
expected arrival time. The red, green and blue histograms are the results from the Monte Carlo
simulation of these events, incorporating known PMT and optical effects. In the red, optical effects
such as reflections and scattering are shut off, so that the time structure is due to effects in
the PMT response, such as resolution and pre- and late-pulsing. In the green and blue histograms, the effects of scattering and reflections (off the wall and the PMT faces) 
are respectively turned on, resulting in photons which actually have delayed arrival times at the PMTs. The right plot shows the charge response of R5912 PMTs in different regions of reconstructed arrival times in laser events.  The observed correlations between the charge and time response
are modeled within the Monte Carlo simulation.

\subsection{Tuning with Electrons} 
\label{sec:omtune}
The results of the external measurements form the foundation for  a model for transporting optical photons through Marcol 7 to their detection by the PMTs in the detector Monte Carlo simulation. The parameters of this model are further tuned using  electrons from cosmic muons decaying-at-rest ($\mu$-DAR) in the detector. These events provide a large sample of easily-identifiable electron events with well-known energy spectrum
that can be reconstructed (as described in Section \ref{sec:rec}) and simulated within the
detector Monte Carlo. The parameters are constrained such that the electrons
from $\mu$-DAR simulated using these parameters  are consistent with those observed in data according to a number of quantities such as the angular and time distribution of the light measured by the PMTs. 

The process starts by randomly drawing a set of parameter values that are consistent with the uncertainties in the parameters. The starting uncertainties are based in large part on the errors
from the  external measurements and physical boundaries.
A sample of electrons  from $\mu$-DAR is simulated with the Monte Carlo using these parameters and reconstructed as in data to determine the position, direction and energy. Based on these reconstructed quantities, the distribution of some target variable
(typically energy, photon arrival times or the geometric distribution of the detected photons) is produced.
This target distribution is compared with the corresponding distribution obtained from the data and 
the $\chi^2$ between the two calculated. 

In each iteration of the procedure, a number of such parameter vectors are drawn, and a sample
of electrons is simulated and reconstructed with each configuration to obtain the target distributions.
Based on the $\chi^2$ probability from the comparison of the target distributions
with the data, a weight is assigned to each vector, resulting in larger weights for parameter vectors for which the target distributions matched the data well, and small weights for those in which the match was poor. The set of weighted vectors are used to produce a new space of allowed vectors that incorporates the constraint of requiring the target distribution to agree with the data, resulting in an effectively smaller space. New vectors can be drawn from this reduced space and the process
is iterated with additional target distributions, higher statistics, or localizing the comparison to
different classes of events. Examples of the latter include a comparison of the energy distribution
in bins of reconstructed radius of the electrons from the center of the detector. 

The determination of the rate of fluorescence and scintillation processes, both characterized by delayed, isotropic light, is particularly
difficult with only electrons, which are relativistic for nearly their entire path length through the detector regardless of energy.
In order to provide an additional handle, neutrino neutral current elastic (NCEL) scattering events are used. These events produce a recoil nucleon in the event with  the outgoing neutrino leaving no trace in
the detector. Since the nucleon is typically below Cherenkov threshold, the NCEL sample 
allows one to study scintillation in an environment where Cherenkov-induced fluorescence is 
suppressed.

The result of the process is a space of optical parameter values in which $\mu$-DAR electrons
simulated with parameter values drawn from the space will have properties which match
those obtained in data. The allowed space is significantly constrained from its initial
configuration based on external measurements. The constrained space
is the foundation  for the estimate of systematic uncertainties due to
detector response in the analysis. Neutrino samples representing both the signal and background
processes are simulated using a number of parameter vectors drawn from the space.
The systematic uncertainties in a given distribution are obtained by considering the
covariance in the distribution across the systematically varied samples. The central values
of the parameters are obtained separately by a more {\em ad hoc} process. In calculating
the covariance for a given distribution, the covariance is taken about the distribution
from the central value, rather than the mean of the distributions obtained from the
systematically varied Monte Carlo samples.


\begin{figure*}[t]
\centering
\includegraphics[width= 82 mm]{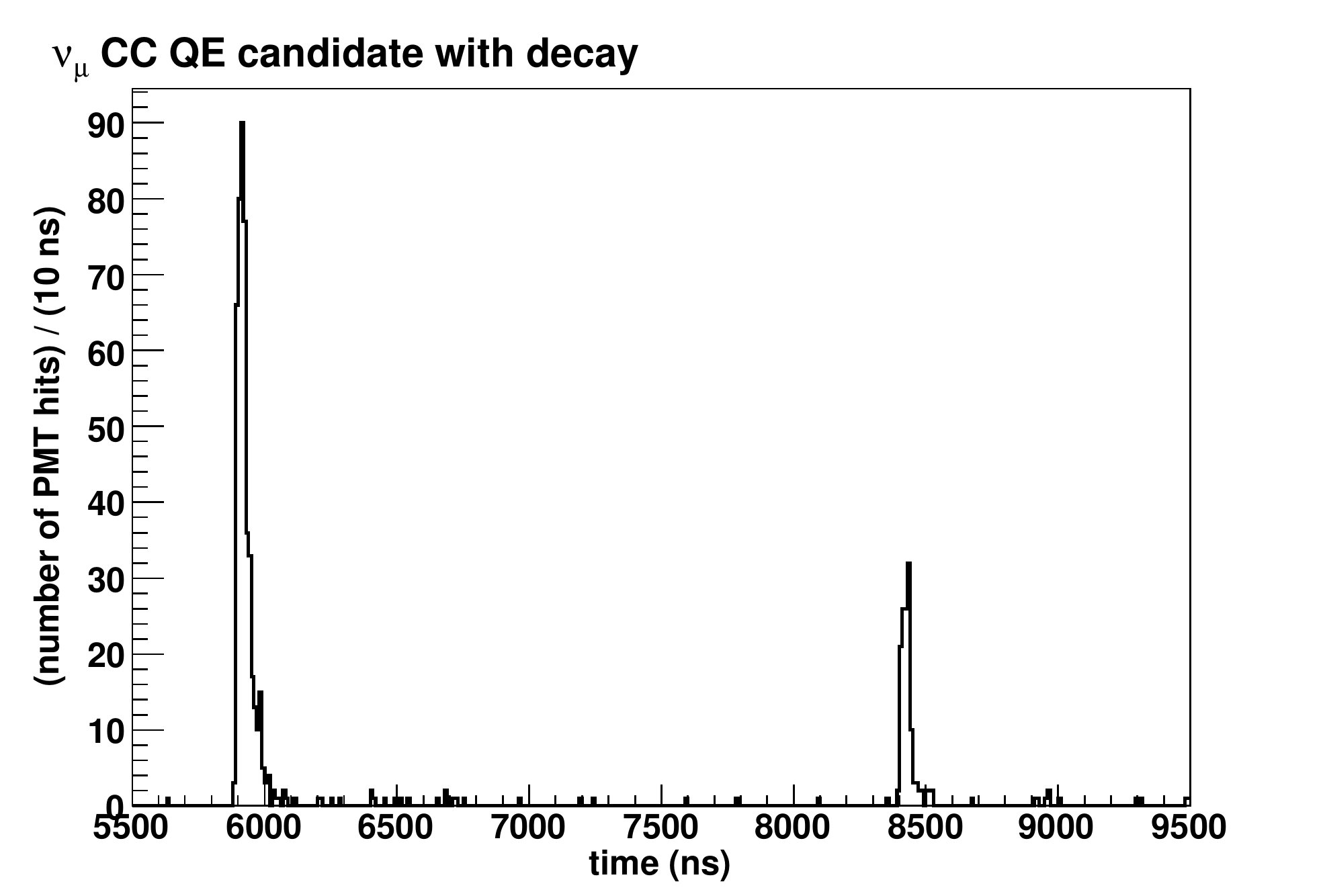}
\includegraphics[width= 78 mm]{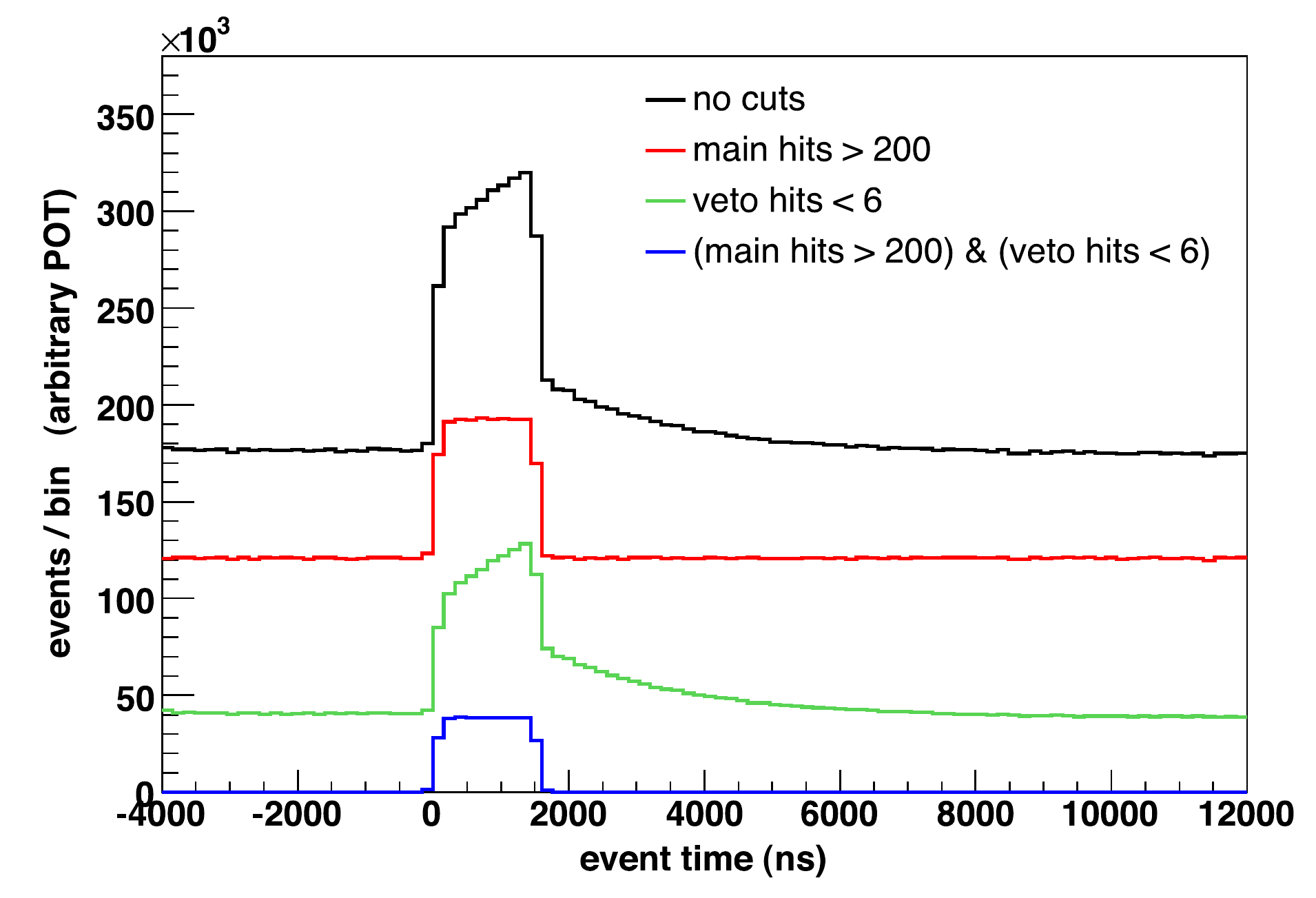}

\caption{Left: Time distribution of PMT hits observed in a neutrino interaction. The first group
of hits at $\sim5.9 \mus$ is consistent with a neutrino interaction in the region occuring
in the beam arrival window of 4.6-6.2 $\mus$. The second, smaller group of hits
at $\sim 8.4\mus$ is consistent with an electron from the decay-at-rest of a muon
produced in the neutrino interaction. Right: Subevent times observed in the beam data.
The  time on the horizontal axis is such that the start of the beam arrival is at 0 ns. The excess
due to beam-induced events can be seen when all subevents are considered (black). The rising
structure within the beam arrival window and the tail after it are due to electrons from the 
decay-at-rest of muons produced in neutrino interactions.
Electrons from $\mu$-DAR, both from cosmic and neutrino-induced muons, are eliminated by requiring $>200$ main hits (red). Cosmic muons are suppressed by requiring
$<6$ veto hits (green). Requiring both $>200$ main hits and $<6$ veto hits results in a neutrino sample with negligible cosmic background (blue), indicated by the absence of subevents occurring outside the
beam arrival window.}\label{fig:subevents}
\end{figure*}

\section{Signal and Background Properties}
At $\mathcal{O}(1 \gev)$ energies, the dominant contribution of charged current (CC) neutrino
interactions is from the charged current quasi-elastic (CCQE) scattering process:
\begin{equation}
\nu_{\ell}+ n \to \ell^- + p
\end{equation}
At MiniBooNE, approximately $40\%$ of all neutrino interactions in the detector are
$\num$ CCQE interactions. The CCQE process is a particularly attractive signal process for
Cherenkov detectors in that the proton is typically below Cherenkov threshold, so that the
light produced in the event is dominated by the outgoing lepton, resulting in a simple event
topology consisting of a single Cherenkov ring. The signal signature is therefore
a Cherenkov ring from the electron emerging from a $\nue$ CCQE interaction.
Furthermore, the quasi two-body kinematics
of the process allow the determination of the incident neutrino energy, up to the initial motion
of the target nucleon in the nucleus, if the momentum of the
outgoing lepton is reconstructed. We refer
to the neutrino energy inferred from the lepton energy and direction under the assumption of CCQE kinematics as $\enuqe$.

The backgrounds come from misidentified $\num$ interactions and the ``intrinsic'' $\nue$
present in the neutrino flux  from muon and kaon decays in the beamline. The dominant
component of misidentified $\num$ events arises from high energy photons,
either from neutral current (NC) $\piz$ production (mainly from neutrino-induced
$\Delta$ resonances) where the $\piz$ decays to
produce two photons, or from $\Delta\to N\gamma$ decays
(also from resonant neutrino scattering). These photons induce electromagnetic
showers much like electrons, resulting in Cherenkov rings which mimic the
signal $\nue$ CCQE process. The NC $\piz$ background can be suppressed to the extent that the second photon from the $\piz$ decay can be detected, while there are no obvious handles for suppressing the radiative $\Delta$ background. Photon backgrounds also arise from
neutrino interactions outside of the detector (``dirt events''). 
The intrinsic background is inherently irreducible,
as it is identical to the signal process in every way apart from their energy spectrum. Constraints
and cross-checks on both sources of background will be considered in Sections \ref{sec:eventrate}
and \ref{sec:xcheck}.

The NUANCE event generator package \cite{nuance} is used to simulate the neutrino
interactions. Based on the predicted flux and the elemental composition of the mineral oil, 
NUANCE will assign an interaction channel and the final state particle configuration. The
effects of interactions within the nucleus, such as the charge exchange and absorption
of pions and the $\Delta+N\to N+N$  process, are simulated according
to a model of the carbon nucleus, as are the emission of photons from the de-excitation
of the nuclear remnant. The neutrino cross sections are tuned to the available data. The
parameters of the CCQE process are adjusted using the observed $Q^2$ distribution in the
the $\num$ CCQE sample described in Section \ref{sec:eventrate}.

\begin{figure*}[t]
\centering
\includegraphics[width= 72 mm]{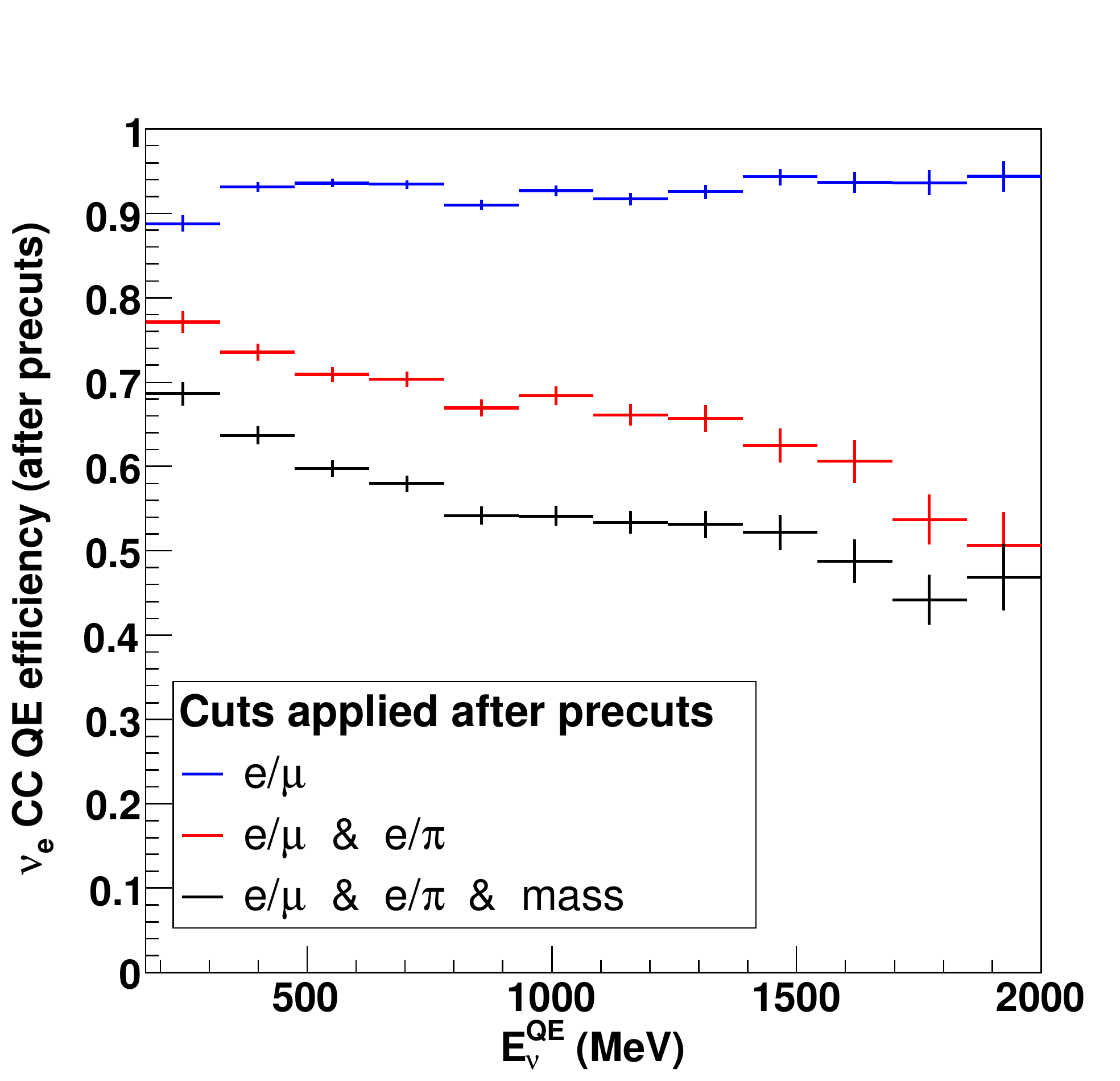}
\includegraphics[width= 97 mm]{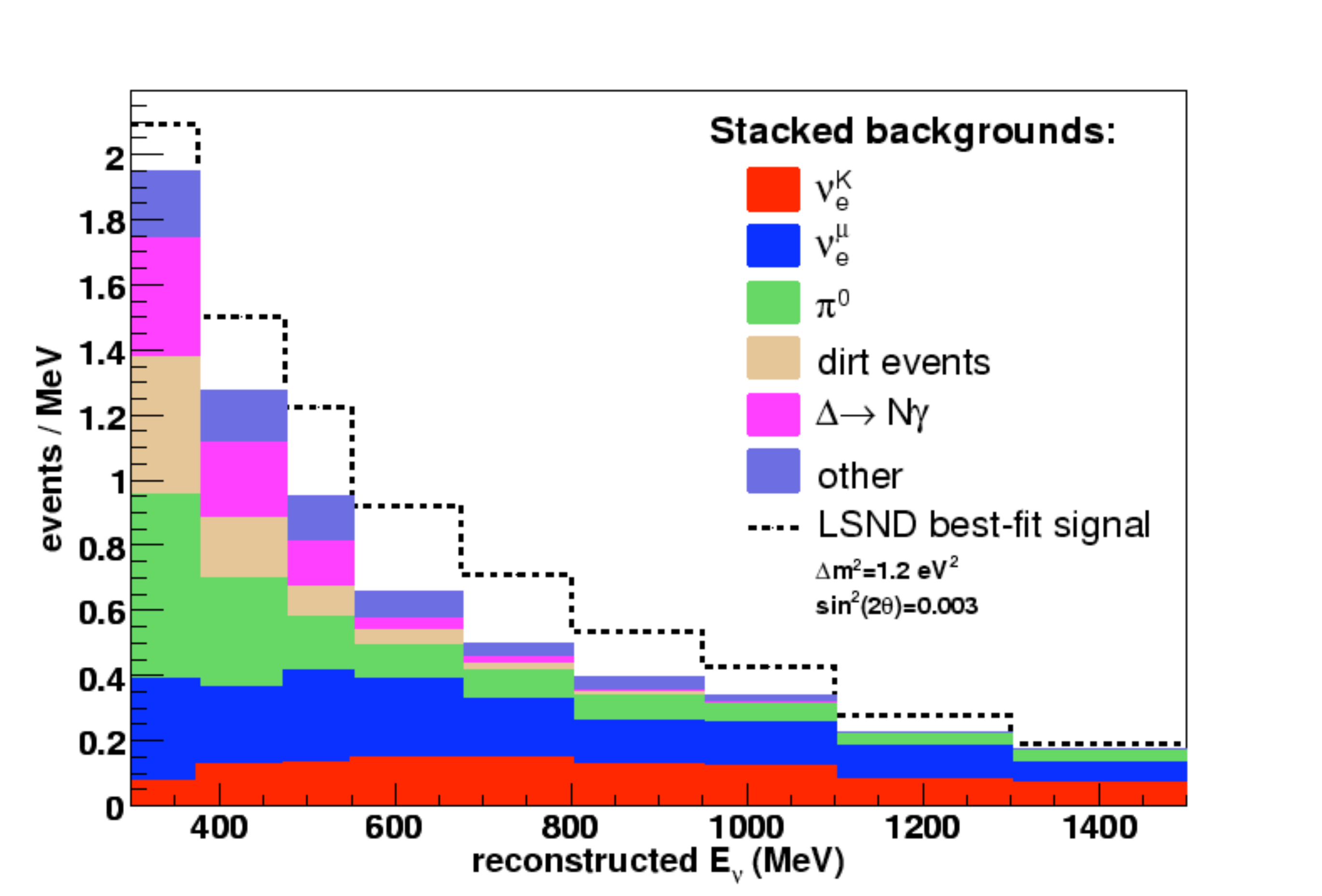}
\caption{Left: Efficiency of the $\nue$ selection as a function of reconstructed neutrino energy following
the preselection. The blue, red and black points show the cumulative efficiency of applying the
$\logemu$, $\logepi$ and $\mgg$ cuts. Right: Expected backgrounds as a function of reconstructed
neutrino energy. The red and blue components show the intrinsic $\nue$ background from kaon
and muon decay, respectively, along with  the various contributions from misidentified $\num$ events
 stacked to give the total background.
The black dashed line shows the total distribution expected in the presence of $\num\to\nue$
oscillations with oscillation parameters corresponding to the best fit values from LSND. } \label{fig:effbkg}
\end{figure*}

\section{Event Reconstruction}
\label{sec:rec}
\subsection{Subevent Identification}
The reconstruction of beam trigger events proceeds by analyzing the time structure
of the PMT hits recorded during the 19.2 $\mus$ window surrounding the
expected arrival time of the neutrino beam. Clusters of hits in time, referred to as subevents,
 that exceed a threshold 10 PMT hits are identified. Neutrino interactions will produce a subevent $4.6-6.2\mus$ after the start of the window, when the beam is expected to arrive. They may produce additional subevents if muons are produced and come to rest within the main region of the detector, leading to decay electrons some microseconds later. The left plot of Figure \ref{fig:subevents} illustrates the time distribution of PMT hits observed in such an event. The first group of PMT hits is consistent in time with a neutrino interaction from the arriving beam, while the second, smaller group of hits is consistent with
an electron from $\mu$-DAR of a muon produced in the neutrino interaction.
These electrons are at lower energy ($<52.8\mev$) than the $\nue$ CCQE
signal events and can be readily distinguished by their low multiplicity of main PMT hits ($<200$).
As illustrated above, these electrons are particularly useful in that they can tag the presence primary muons produced in $\num$ CC interactions, as well as $\pip$ which undergo the $\pip\to\mup\to\ep$ decay chain.

A subevent can be identified as either an incoming cosmic muon or an uncontained neutrino
interaction by the number of veto hits it contains: a single penetration through the veto by
a muon typically leaves 18 veto hits, while a through-going muon that enters and exits
the detector will typically leave $>25$ veto hits. With a simple combination of requirements
on the number of main and veto hits, subevents due to cosmic activity  can be  eliminated, leaving a pure neutrino sample as illustrated in the right plot of Figure \ref{fig:subevents}.

\subsection{Track Reconstruction}
Following the identification of subevents, the PMT hits associated with a subevent
are fit under four hypothesized track configurations for the the outgoing particles in the event.
The first pair of fits assume a single track, either an electron or a muon, emerging from a  certain point and time (four parameters) in the detector, in a certain direction and with a certain kinetic energy (three parameters). Based on a vector of parameter values and a track hypothesis, a predictive model determines the expected number and arrival times of photoelectrons at each of the main PMTs. Convoluting these predicted values with a model of the PMT response, the actual charge and time of the hits in the subevent (including PMTs which did not register hits) can be compared to the model prediction to calculate a likelihood.
Using the standard method of maximum likelihood,  the best-fit parameter values are extracted by
varying the seven parameters in such a way as to maximize the likelihood using 
MINUIT\cite{minuit}. We note that the single
electron track model  corresponds to the signal $\nue$ CCQE hypothesis, while the single
muon track model corresponds to the the $\num$ CCQE hypothesis. Both fit models assume that the
outgoing lepton is the dominant source of light in the event, with negligible contribution from the
recoil hadron system.

A second set of fits involve a twelve parameter model with two electron tracks. In this model,
two tracks emerge from positions displaced from the vertex, with each track pointing
 back to the vertex. This models the $\piz\to\gamma\gamma$ decay, with the displacement of
 each of the tracks from the vertex accounting for the distance traversed by each photon before
 showering. The model assumes
 that the detector response to each photon is the same as an electron at the  conversion point with the same time, direction and energy. The
 twelve parameters correspond to the position and time of the vertex (four parameters), and the direction, energy, and displacement (four parameters) of each track. 
 
 A second two-track (2T) fit involves the same model with the kinematics of the two tracks confined in such a way that the invariant mass is equal to the nominal $\piz$ mass, resulting in a model with eleven free parameters.
 The 2T fits, which use the results of the one-track electron fit as a starting point, proceed conceptually
 in the same way as the one-track fit in that the model produces the predicted time and charge
 at each PMT in the subevent based on the configuration of the two tracks, and employ
 the maximum likelihood method to extract the best-fit parameter values.  Practically,
 the 2T fits differ in  that a number of different configurations are tested for each subevent, with the most promising configurations used as starting configurations for the likelihood maximization process in MINUIT.
 The results of the fits are a complete kinematic reconstruction of the $\piz\to\gamma\gamma$ decay,
 with and without a mass constraint. In particular, with the momentum of each photon reconstructed,
 one  can infer the invariant mass of the $\gamma\gamma$ system (trivial in the
case of the fixed-mass 2T fit) and the momentum of the $\piz$.

\section{Event Selection}
The selection of candidate $\nue$ CCQE events from the beam data starts with a set of simple requirements to eliminate obvious $\num$ CC events (only one subevent identified in the event), incoming cosmic ray muons (less than six veto hits in the subevent) and electrons from $\mu$-DAR (greater than two hundred main hits in the subevent). The average time of main PMT hits in the subevent
is required to be consistent with the expected beam arrival window.

This selection identifies a sample of neutrino candidates with neglible contamination from
cosmic backgrounds that then undergo track reconstruction.
Based on the electron fit, the reconstructed vertex of the event is required to be within
500 cm of the center of the detector, where the the radius defined by the face of the
main PMTs is 548.5 cm. The results of the muon fit are used to determine the projected endpoint
of the track assuming that it is a muon, which is required to be within 488 cm of the center of the detector. This
eliminates $\num$ CC events in which the muon decays close to or behind the main PMT array, resulting in a decay electron that may evade detection. The reconstructed energy
from the electron fit is required to be greater than $140\mev$, while the reconstructed neutrino energy 
$(\enuqe)$ is required to be in the range $475-3000\mev$. 

A number of criteria are imposed to reduce the background from $\num$ interactions.
A cut on $\log({\mathcal L}_{e}/{\mathcal L}_{\mu})$, the logarithm of the ratio of the likelihoods returned
from the  electron and muon fits, reduces the remaining $\num$ CC background. This quantity
tests whether the event fit better under the electron track hypothesis or the muon-track
hypothesis. Similarly, a cut on $\log({\mathcal L}_e/{\mathcal L_\pi})$, where ${\mathcal L}_\pi$ is the
likelihood returned by the fixed-mass 2T fit, rejects events which fit better
under the $\piz$ hypothesis than the electron hypothesis. Finally, a cut on events which have a large
reconstructed mass from the free-mass 2T fit are also rejected. In all three cases,
the specific cut values are a function of the reconstructed energy from the electron
fit and optimized simultaneously using Monte Carlo-simulated samples to maximize the signal sensitivity. 

A blind analysis is implemented by sequestering from study any data events which
pass all the selection criteria (including the number of such events) until all aspects of the analysis are finalized. The unblinding procedure is described in Section \ref{sec:unblinding}. In the process
of performing cross checks on background rates, two subsamples of the signal events which
are dominated by particular backgrounds were unblinded as described in Section \ref{sec:xcheck}. 

Figure \ref{fig:effbkg} shows the signal efficiency for $\nue$ CCQE events for the three $\num$ cuts
as a function of $\enuqe$ and the expected background distribution in
$\enuqe$.  The denominator for the efficiency plot is all events in the
$\enuqe$ bin satisfying the main and veto hit cuts as well as the fiducial  requirements. The background plot is normalized to show the expected background rate for the $5.58\times 10^{20}$ protons-on-target used in this analysis. This plot also shows the expected signal distribution for the LSND best-fit oscillation parameters. Table \ref{tab:backgrounds} summarizes the expected
signal and background rates for events with  $475<\enuqe<1250\mev$, where most of a potential
oscillation signal is expected to lie.

\begin{table}[h]
\begin{center}
\caption{Expected background rates and systematic uncertainties for the $\nue$ CCQE selection 
in the energy range $475 < \enuqe < 1250 \mev$. The expected signal yield for
a $0.26\%$ $\num\to\nue$ transmutation is also shown. The systematic uncertainties
are correlated across different background sources in some cases.}
\begin{tabular}{lr}\hline
\textbf{Process}				& \textbf{Number of Events} \\ \hline
$\nu_\mu$ CCQE				&$10 \pm 2$ 		\\
$\nu_\mu e \rightarrow \nu_\mu e$	& $7 \pm 2$            \\
Miscellaneous $\nu_\mu$ Events	&$13 \pm 5$	 	 \\
NC $\pi^0$					&$62 \pm 10 $            \\
NC $\Delta \rightarrow N \gamma$	&$20 \pm 4$		 \\
NC Coherent \& Radiative $\gamma$&$<1$	 \\
Dirt Events					&$17 \pm 3$  \\ \hline
$\nu_e$ from $\mu$ Decay		&$132 \pm 10$ \\
$\nu_e$ from $K^+$ Decay		&$71 \pm 26$ \\
$\nu_e$ from $K^0_L$ Decay		&$23 \pm 7$ \\
$\nu_e$ from $\pi$ Decay		&$3 \pm 1$ \\ \hline
Total Background 				&$358 \pm 35$ \\ \hline
0.26\% $\nu_\mu \rightarrow \nu_e$&$163 \pm 21$ \\ \hline
\end{tabular}
\label{tab:backgrounds}
\end{center}
\end{table}

\section{Boosted Decision Tree Selection}
A second event selection system for signal $\nue$ CCQE candidates based
on a boosted decision tree (BDT) \cite{boost} is used to cross check the likelihood ratio-based
selection. This method uses a simpler event reconstruction with corresponding
electron, muon and two-track hypotheses.
 Based on this fit, 172 variables characterizing the space and time distribution 
of the light are formed using quantities like the charge and time likelihoods
in different angular regions, the likelihood ratios formed from the three fits,
and the parameters from the two-track fits. These quantities form the input
for the BDT, which is trained using a ``cascade'' technique \cite{cascade} on 
simulated samples of background and signal events. For each event,  the BDT polls
the output of its decision trees to determine the final output. 

\section{Background Measurements}
\label{sec:eventrate}
The rate of a number of background processes can be measured or inferred from the neutrino data collected at MiniBooNE. One such measurement is based on the
identified  $\num$ CCQE events which provide a measure of the dominant
$\num$ flux. A second measurement is  the rate of NC $\piz$ events. In both cases,
the direct determination of the rate within the data allows one to reduce systematic uncertainties
associated with the Monte Carlo-based prediction of the neutrino flux and the neutrino cross sections.

\begin{figure*}[t]
\centering
\includegraphics[width= 70.0 mm]{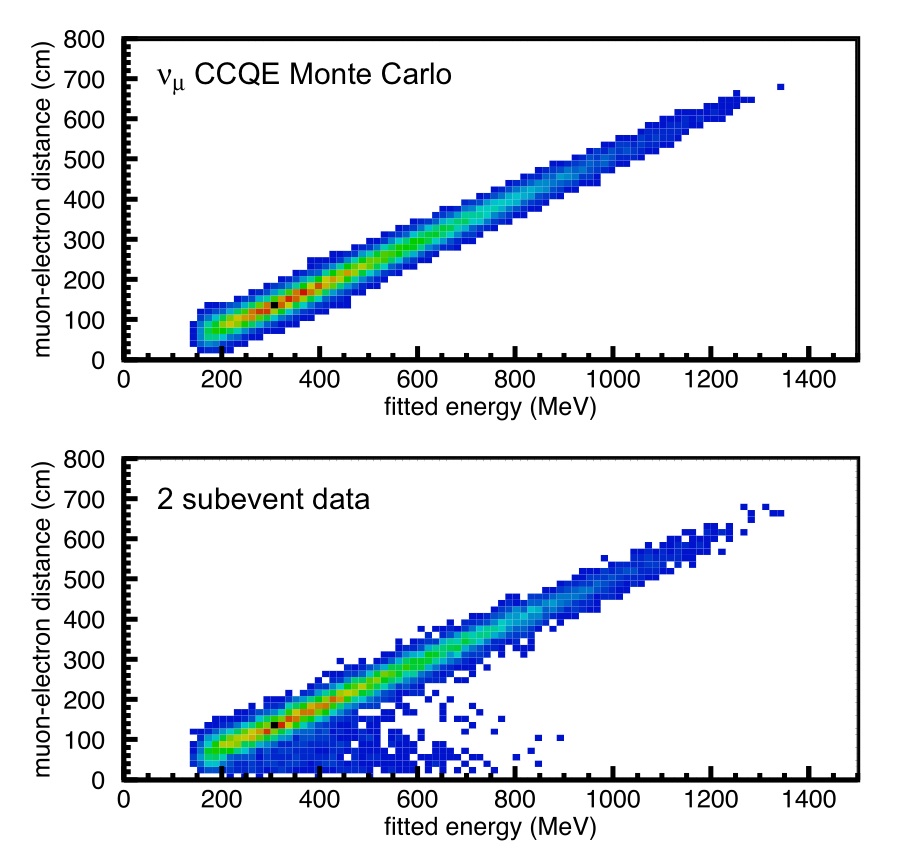}
\includegraphics[width= 90.0 mm]{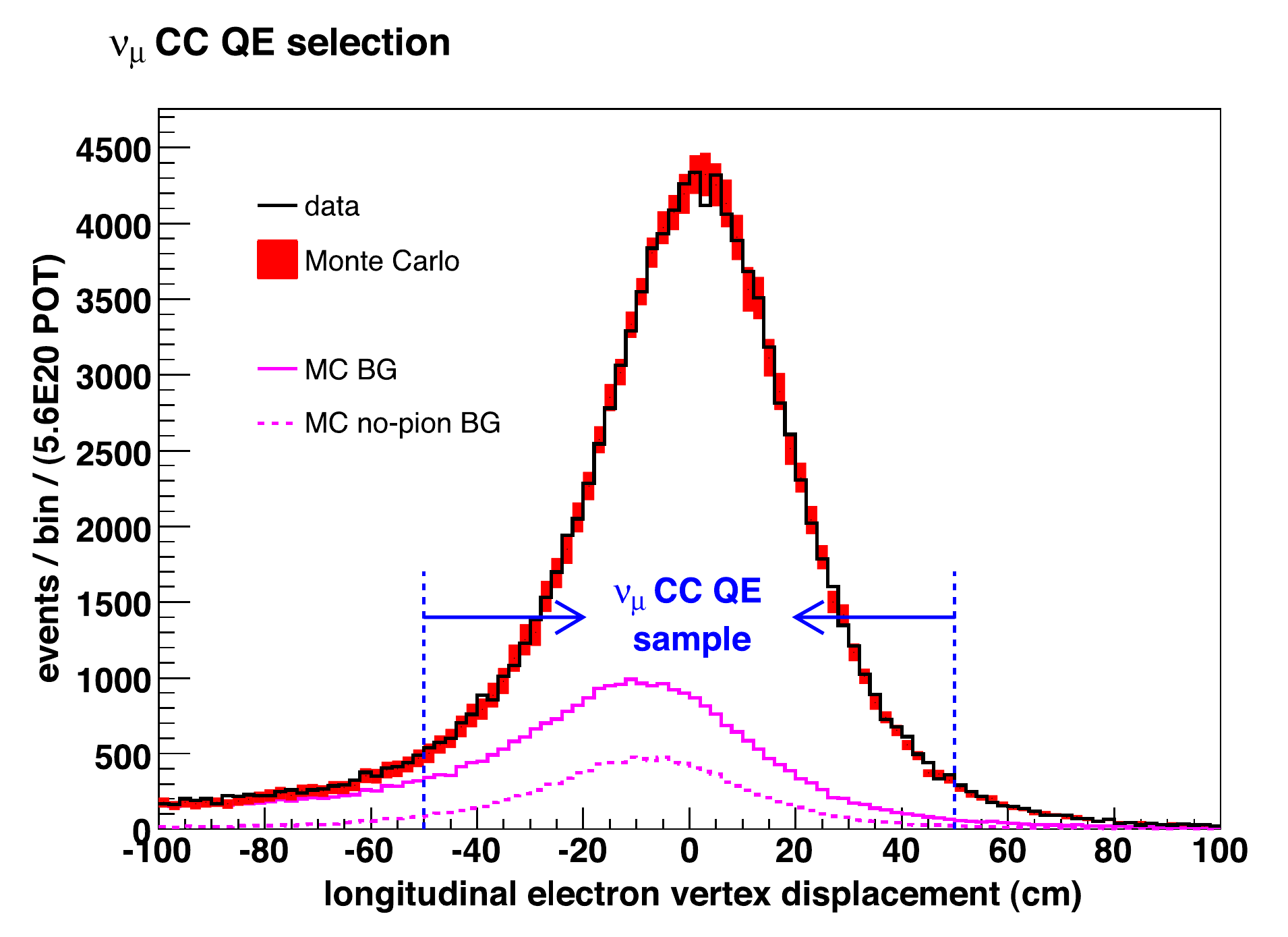}
\caption{Left: Reconstructed muon/electron distance versus visible energy for a sample
Monte Carlo-simulated $\num$ CCQE events (top) and data (bottom) where each sample
is required to have two subevents consistent with a neutrino interaction and an electron
from muon decay-at-rest. Right: the longitudinal displacement of the reconstructed electron
vertex relative to extrapolated muon endpoint as described in the text. The non-$\num$ CCQE
backgrounds are shown as the solid pink histogram. The dashed pink shows the subset of the
background which contains no pions in the final state.} \label{fig:mumichel}
\end{figure*}

\subsection{$\num$ CCQE events}
The selection of $\num$ CCQE events starts by identifying events
with a single subevent consistent with an electron from $\mu$-DAR following the primary subevent.
The primary subevent is  required to occur in the beam arrival window and have less than
six veto hits and greater than two hundred main hits in order to suppress cosmic background.
The second subevent is required to have less than six veto hits and less than two hundred main hits,
rejecting cosmic muons  and ensuring that the energy is consistent with an electron from $\mu$-DAR.  Following the reconstruction of both subevents with the single-track reconstruction, the geometry of the two subevents is analyzed to determine whether the electron vertex is consistent
with the endpoint of the muon.

This consistency is determined by comparing the muon track length determined
from the reconstructed muon and electron vertices, and the expected muon track length
based on the reconstructed muon energy.
For  $\num$ CCQE events, where the muon is the dominant source of light, this is expected to be linear relationship due to the minimum-ionizing behavior of the muon. This is demonstrated in the top left figure of Figure \ref{fig:mumichel}, which shows the reconstructed track length versus muon kinetic energy for a sample of Monte Carlo-simulated $\num$ CCQE events. The bottom figure shows the same distribution observed
in the data. The core of the distribution lies on a ``line'' with slope $\sim\! 2\mev/\cm$, but a cluster of
events lies below it, corresponding to background events where extra particles in the hadronic system disrupt the relationship. These particles introduce more energy into the event while
the reconstructed track lengths, based on vertex information, are not affected. This places the
events below the line (more energy per track length).

A precise form for this line is determined by profiling the two dimensional distribution observed in data and fitting it to a linear function. By using the data directly (rather than Monte Carlo-simulated
$\num$ CCQE events), systematic uncertainties associated with energy scale are reduced. The
function summarizes the predicted energy/track length relationship for  $\num$ CCQE events and is used by comparing the distance between the projected
muon endpoint (using the reconstructed energy, direction and the energy/track length relationship)
and the reconstructed electron vertex. The vector difference between these two points is 
projected along the reconstructed muon direction resulting in the longitudinal electron vertex displacement. This distribution of this quantity for two subevent $\num$ CC candidates is shown on the right of Figure \ref{fig:mumichel}. The background, particularly those with pions in the final state, lie
towards negative values (electron vertex closer than would be expected based on the reconstructed
muon energy) while the $\num$ CCQE events are centered at zero. The $\num$ CCQE candidates
are selected by requiring that the longitudinal electron vertex displacement is less than $50\cm$.

The $\enuqe$ distribution  of the selected $\num$ CCQE events is used in two
ways to determine event rates of other processes. 
The first is  the rate and spectrum of signal $\nue$ oscillation events
for a given set of oscillation parameters ($\sin^2 2\theta$, $\Delta m^2$). For small
oscillation probabilities, the rate of $\num$ CCQE events can be translated 
into the rate of oscillation events since they originate in the same flux of neutrinos
and interact via the same CCQE process. Differences between $\num$ and $\nue$
cross sections are considered in the systematic uncertainties.
In order to make this translation, one must account for the effects of resolution and efficiencies. The detector Monte Carlo simulation is used to infer the energy spectrum of the $\num$ events from the observed  $\enuqe$ distribution, which is
then used to determine the rate of oscillation events as a function of neutrino energy.

This determination reduces systematic
uncertainties in the predicted signal yield by using the observed rate $\num$ CCQE rate directly
to infer the rate of  $\nue$ CCQE events. 
The procedure is also used to correct the rate of other $\num$-induced processes (apart from
NC $\piz$ and radiative $\Delta$ events), though the  systematic uncertainties
are typically larger when trying to relate the rate of processes which interact with intrinsically different
neutrino cross sections.

Second, the dominant source of the intrinsic $\nue$ from muon decay (the largest single background
source) are from decays of the same $\pip$ which decay to produce the
$\num$ observed in the $\num$ CCQE sample. As a result,  the rate of muons from $\pip$ decays
in the beamline can be inferred
from the observed rate of these events in the detector. Furthermore, due to the tight kinematic
constraints for  $\pip\to\mup+\num$ decays to send a neutrinos towards the detector
({\em i.e.} only decays with forward $\num$ are observed), there is a close relationship
between the observed $\num$ energy spectrum and the energy spectrum of the parent pions.
This allows the spectrum of this background to be inferred from the neutrino
data. Residual systematic uncertainties result from the detector response uncertainties and
neutrino cross section uncertainties which affect the efficiency  of the
$\num$ CCQE selection and the background rates. Since the $\num$ and $\nue$ events which result from pions of a given 
energy have different energy spectra (the former are from the two-body decay of the $\pip$,
while the latter are from the three-body decay of $\mup$ from the two-body decay of $\pip$),
there are additional uncertainties associated with the energy dependence of the 
neutrino cross sections.

\begin{figure*}[t]
\centering
\includegraphics[width= 150.0 mm]{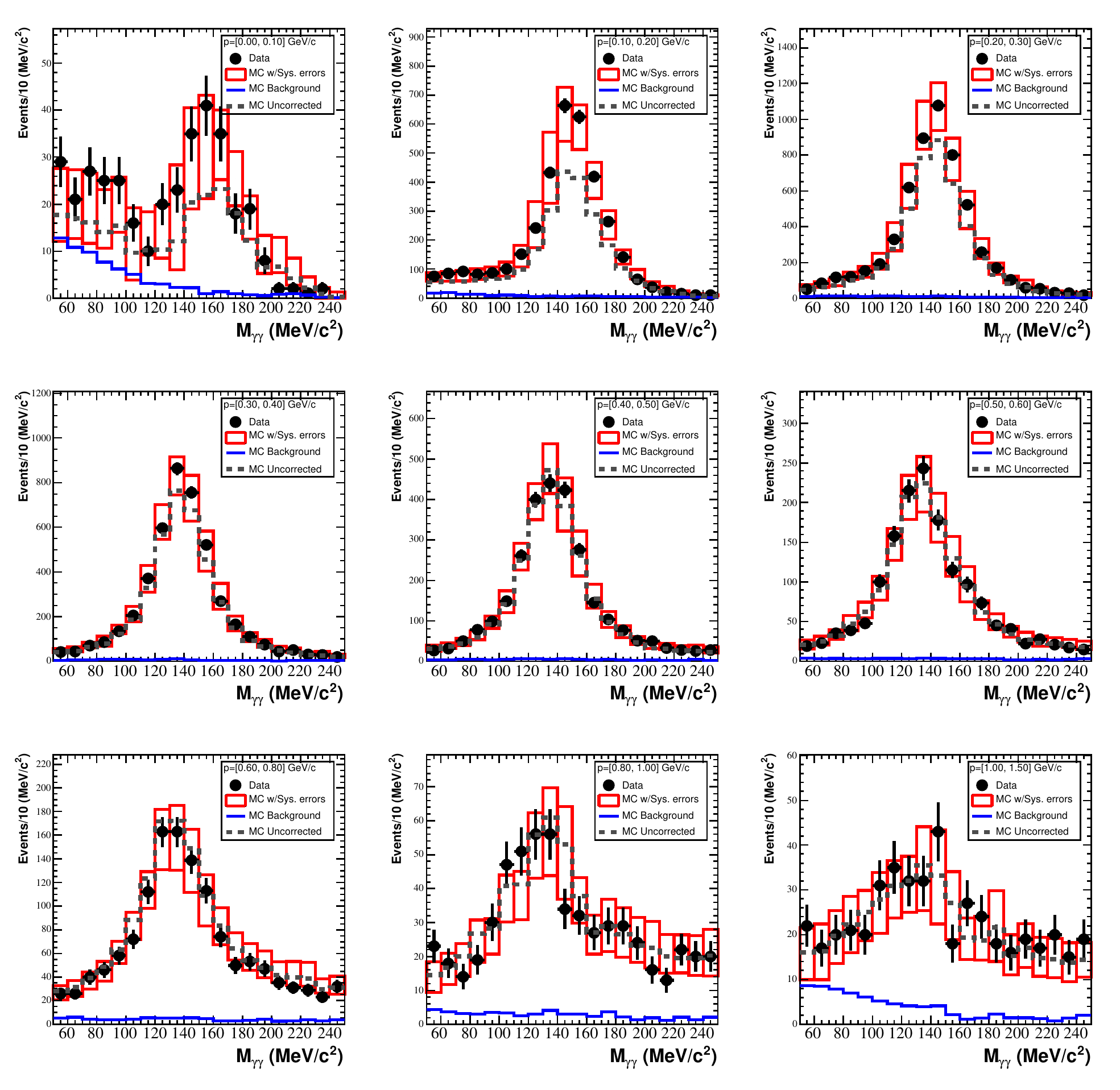}
\caption{Observed $\mgg$ distributions in the $\piz$ control sample in bins
of reconstructed $\piz$ momentum, from lowest (top left) to highest (bottom right).
The black points show the observed data. The blue histograms show the expected contribution from non-$\piz$ events, while the red histogram shows the total Monte Carlo prediction with systematic errors following the correction procedure described in the text. The dashed grey line shows the Monte Carlo prediction prior to the correction. } \label{fig:pi0}
\end{figure*}

\subsection{NC $\piz$ events}
A large fraction of the  NC $\piz$ events are readily identifiable by the invariant mass determined
by the free-mass 2T fit. The events in the $\piz$ peak of the invariant mass distribution are used
as a control sample in which the rate and kinematic properties of the NC $\piz$ events are studied.
This rate is used to correct the rate and spectrum of the NC $\piz$ background contribution to 
the signal $\nue$ CCQE  sample. While the extrapolation between these disjoint sample depends on the detector Monte Carlo simulation to predict  the efficiency of selecting NC $\piz$ events, the 
rate of non-NC $\piz$ events within the control sample, and the fraction of NC $\piz$ events that
pass the $\nue$ selection criteria, the systematic uncertainties incurred by a rate estimation using
the predicted  neutrino flux and cross sections are greatly reduced by measuring the production rate directly in the data.

The event selection for the control sample starts by requiring a single subevent with greater than 200 hits in the main PMT array and less than 6 veto hits. The single subevent requirement is 
used since NC $\piz$ events are not expected to have muons. The remaining
$\num$ CC background is suppressed by requiring $\logemu>0.05$.
Electron-like events are eliminated by requiring $\logepi<0$
(likelihood ratio favoring $\piz$ rather than $e$) and $90 < m_{\gamma\gamma}<180\mevcc$
(reconstructed mass consistent with nominal $\piz$ mass). The selection results in a NC $\piz$
sample of very high purity (typically $>95\%)$ so that systematic uncertainties due to 
non-NC $\piz$ events are minimized.

The selected NC $\piz$ candidate events are divided according to their reconstructed momentum into nine bins ranging from 0 to $1500\mevc$, as shown in Figure \ref{fig:pi0}.
Non-NC $\piz$ backgrounds are subtracted according to the predicted rates from the Monte Carlo
simulation.  The selection efficiency and momentum resolution predicted by the Monte Carlo simulation are used to infer the absolute rate of NC $\piz$ events in true momentum bins prior to the
selection. The NC $\piz$ rates in the Monte Carlo simulation are corrected so 
that the predicted rates in the control sample match the observed rate.
The same momentum-dependent corrections are applied to the NC $\piz$ events that
enter as background into the signal $\nue$ CCQE sample.

The systematic uncertainties in the $\piz$ rate analysis are dominated by the uncertainties in the detector response and the composition of the NC $\piz$ events. The latter uncertainty is associated with the fact that the analysis is performed agnostically to which exclusive channel (resonant, coherent, etc.)
is producing the $\piz$. On the other hand, the details of the final state, such as the  presence of other pions,  the energy of the recoil nucleon, etc., affect the efficiency of the selection and the reconstructed
kinematics. As a result, the extracted yield of NC $\piz$ events is sensitive to the underlying
composition of exclusive channels. 

The measured rate of NC $\piz$ production also constrains the rate of the
$\Delta\to N\gamma$ process, since the dominant source of NC $\piz$
production is $\Delta$ production and the two processes are related by the
electromagnetic and hadronic branching ratios of the $\Delta$. The predicted rate of
$\Delta\to N \gamma$ events is corrected according to the observed NC $\piz$ rate. This
correction is also subject to the uncertainties in the channel composition of the observed
$\piz$ events, most notably the uncertainty in the fraction of NC $\piz$ events that originate 
from $\Delta$ production.

\begin{figure*}[t]
\centering
\includegraphics[width= 80.0 mm]{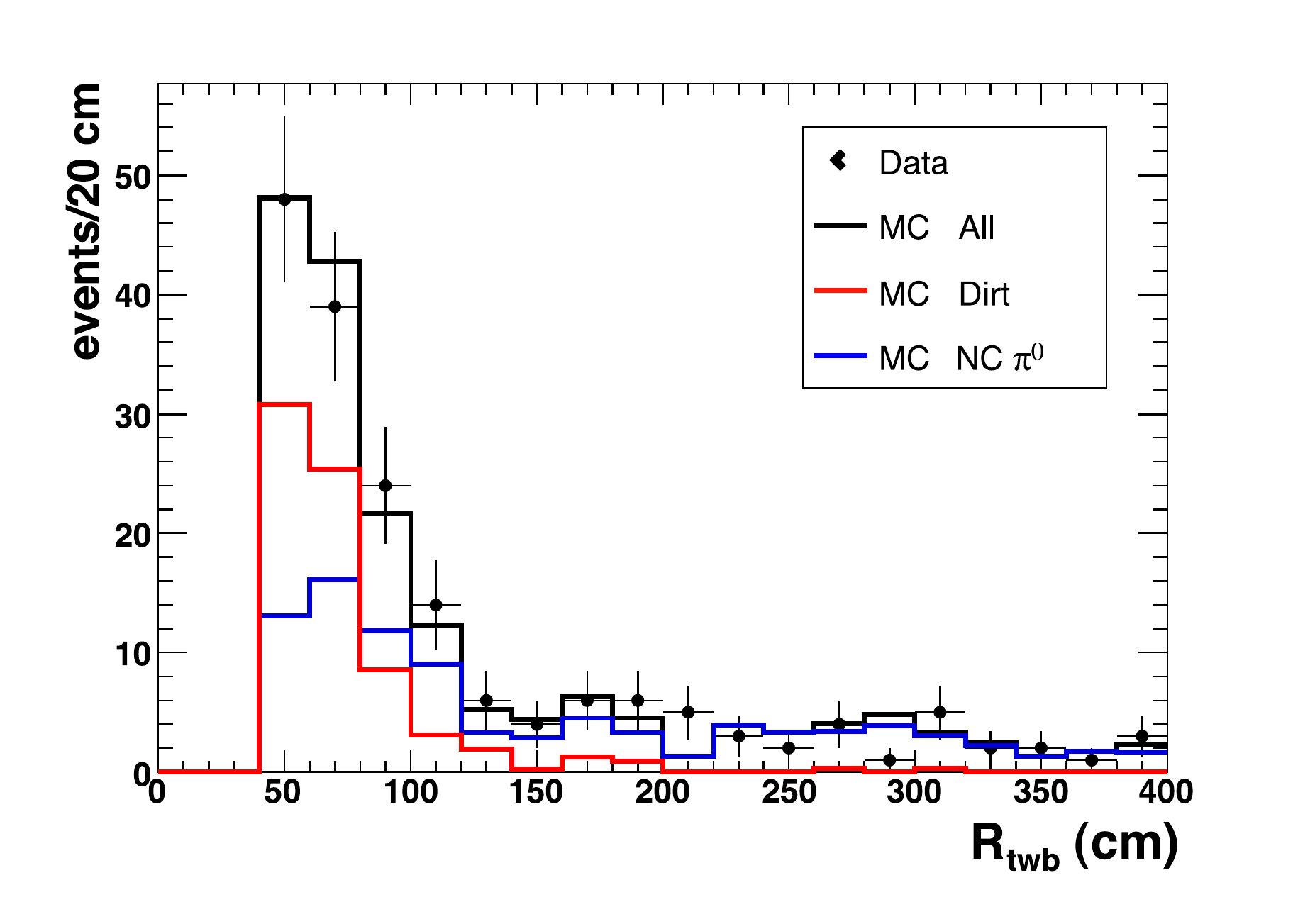}
\includegraphics[width= 80.0 mm]{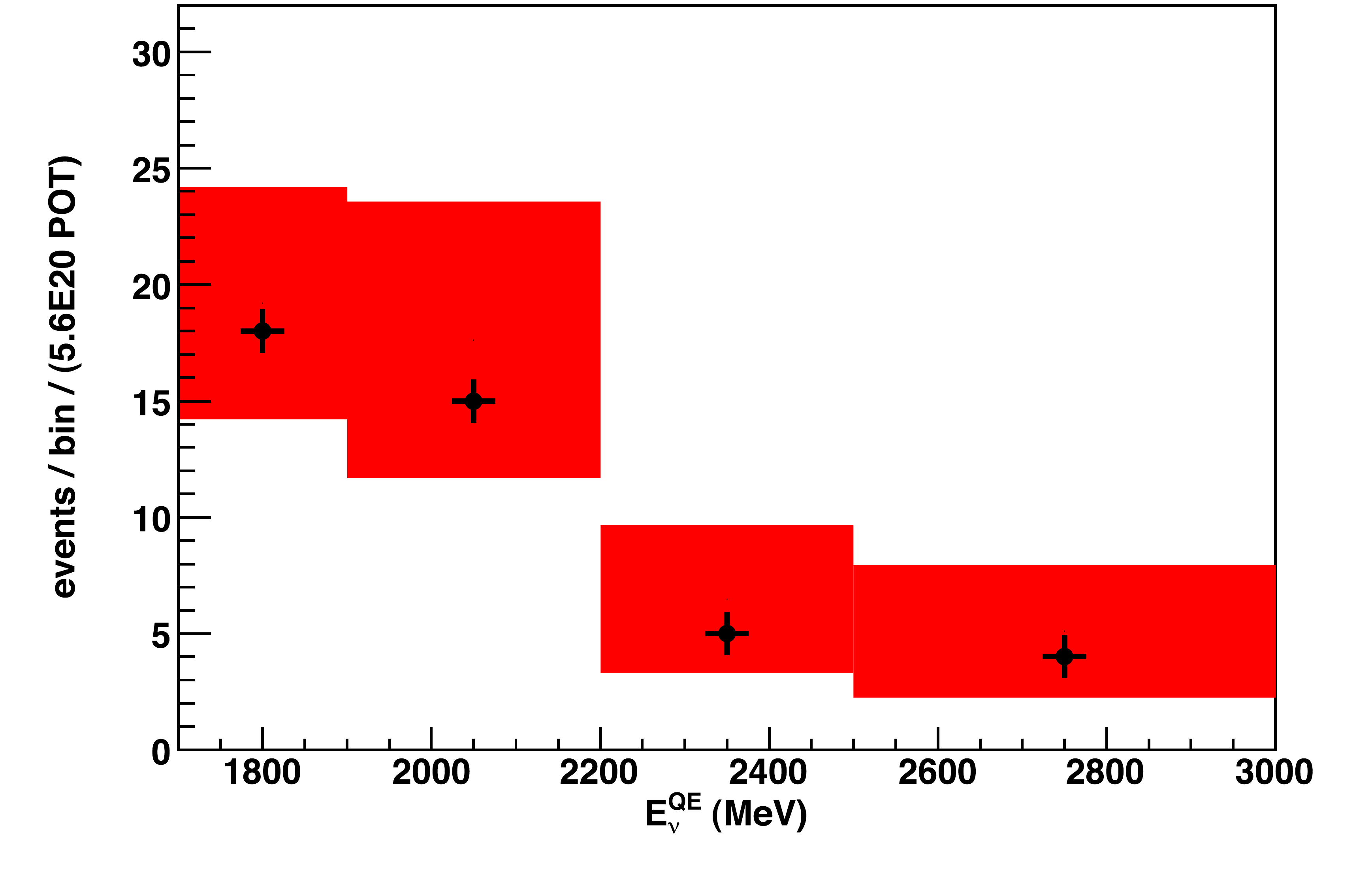}
\caption{Left: The $R_{TWB}$ (distance of the vertex from the wall in the direction
opposite to the track direction) distribution for low energy $\nue$ candidate events. Right:
Reconstructed neutrino energy ($\enuqe$) distribution for events with $\enuqe>1600\mev$. The data are shown as black points while the Monte Carlo-predicted background is shown in red, where the
boxes indicate the uncertainties. Both distributions are absolutely normalized to $5.58\times 10^{20}$ protons-on-target.} 
\label{fig:xcheck}
\vskip 0.5 cm
\end{figure*}

\section{Background Cross Checks}
\label{sec:xcheck}
In order to check the Monte Carlo-based background predictions for the signal $\nue$ candidate
sample, a number of control samples and distributions in the data are investigated. The control
samples are constructed to enrich a particular type of background. The distribution of geometric,
kinematic and background-suppression variables are compared between data and the Monte Carlo
simulation via  a $\chi^2$ test including all sources of error and their correlations to check for
consistency. We describe a few of these important cross checks here.

The behavior of misidentified NC $\piz$ events, the dominant source of $\num$ background,
was examined using sideband regions in the two background suppression variables $\logepi$
and $\mgg$. While the overall rate of these events has been corrected to match the observed
rate within the control sample of well-reconstructed NC $\piz$ events, these sidebands cross check
the behavior of the background with events which are signal-like in one of the variable (signal-like
in both variables would result in events which are in the blinded signal region).  The rate
and kinematic properties of the events, as well as the distribution of the background suppression variables, are found to be consistent between the Monte Carlo simulation, indicating that this
background is properly modeled.

A second source of $\num$ background comes from neutrino interactions outside of the detector which 
produce high energy photons (typically from $\piz$ decay) referred to as ``dirt'' events. The
veto region is approximately one radiation length in thickness. This results in some of these
photons evading detection in the veto and showering within the main region, yielding electron-like Cherenkov rings. They are characterized by their proximity to the edge of the detector and their relatively
low energy (typically 200-300$\mev$). A control sample enriched in these events is isolated
by considering events which otherwise pass the $\nue$ selection criteria, but have a reconstructed vertex close to the wall and low visible energy. To further isolate the contribution from these events, one examines the distribution of events in the variable $R_{TWB}$, which considers the distance of the vertex from a  540-cm-radius sphere concentric with the detector along the direction opposite to the reconstructed track direction.
Since dirt events are inward-directed and near the edge of the main region, they 
concentrate at low $R_{TWB}$. The distribution of $R_{TWB}$ in this control sample is shown on
the left in Figure \ref{fig:xcheck}. The predicted contributions in the Monte Carlo simulation
for the two major components of events  in this sample, namely NC $\piz$
production within the detector and the dirt events, are also shown. The distribution terminates at 40 cm due
to the requirement that the event vertex lie within 500 cm of the center of the detector. 
 The consistency between the Monte Carlo simulation and the data indicate that the rate of this
 background is accurately modeled in the simulation. 

Events at high energy ($\enuqe>1.6\gev$) have a large contribution from intrinsic $\nue$  from kaon decay. The expected
rate of $\nue$ events from oscillations is also small in this energy region, as seen in the right
plot of Figure \ref{fig:effbkg} .The $\enuqe$ spectrum of $\nue$ candidates passing the selection criteria with $\enuqe>1.6\gev$ is shown on the right in Figure \ref{fig:xcheck}. The yield of events in this control sample is consistent with the default background prediction, though the systematic uncertainties, primarily due to the uncertainty in the rate of high momentum $\piz$ production, are large.
The $\nue$ candidates events in this energy range are incorporated in to the signal extraction
procedure described in Section \ref{sec:unblinding}, thus constraining the rate of intrinsic background.

\begin{figure*}[t]
\centering
\includegraphics[width= 70.0 mm]{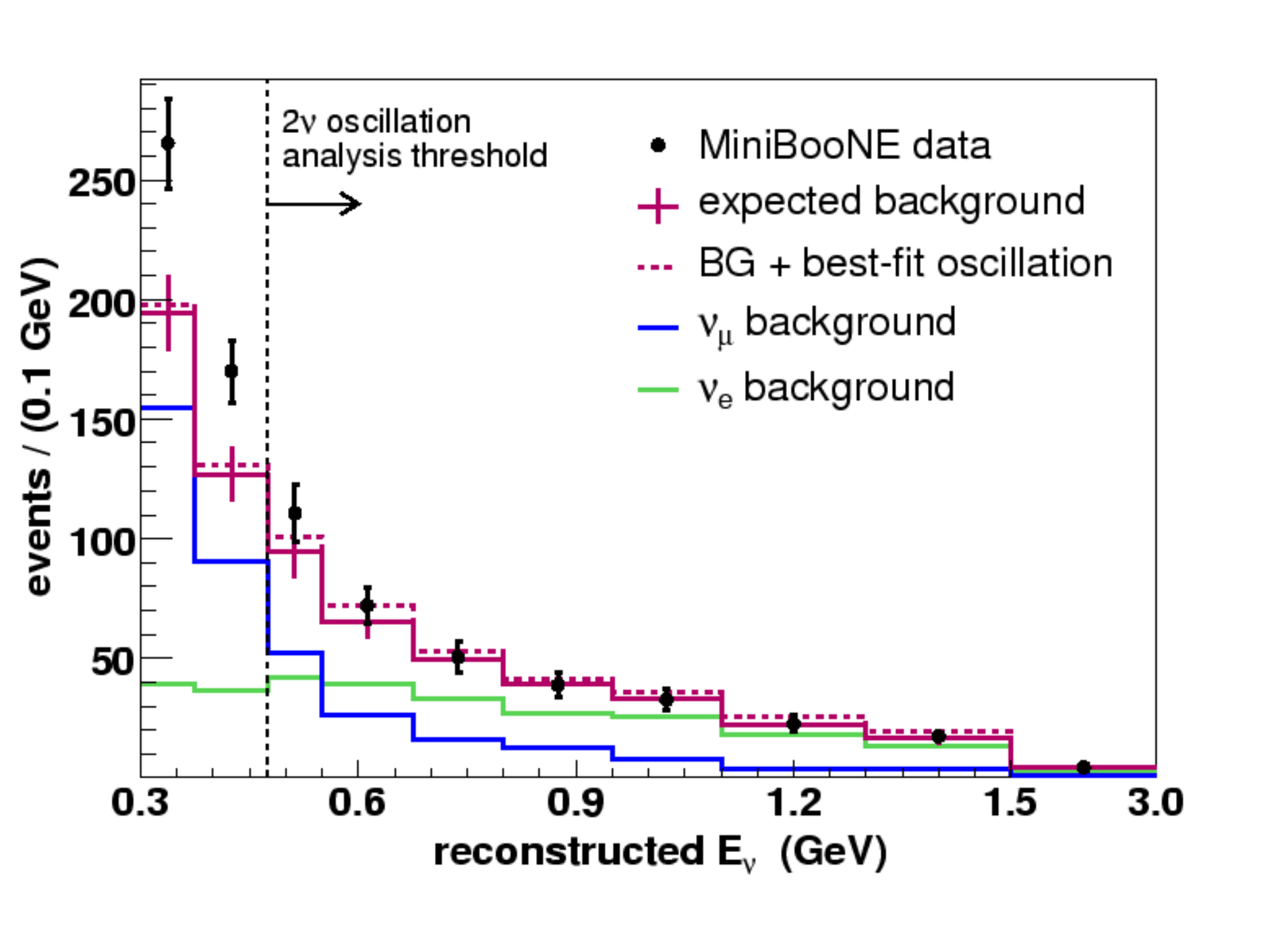}
\includegraphics[width= 90.0 mm]{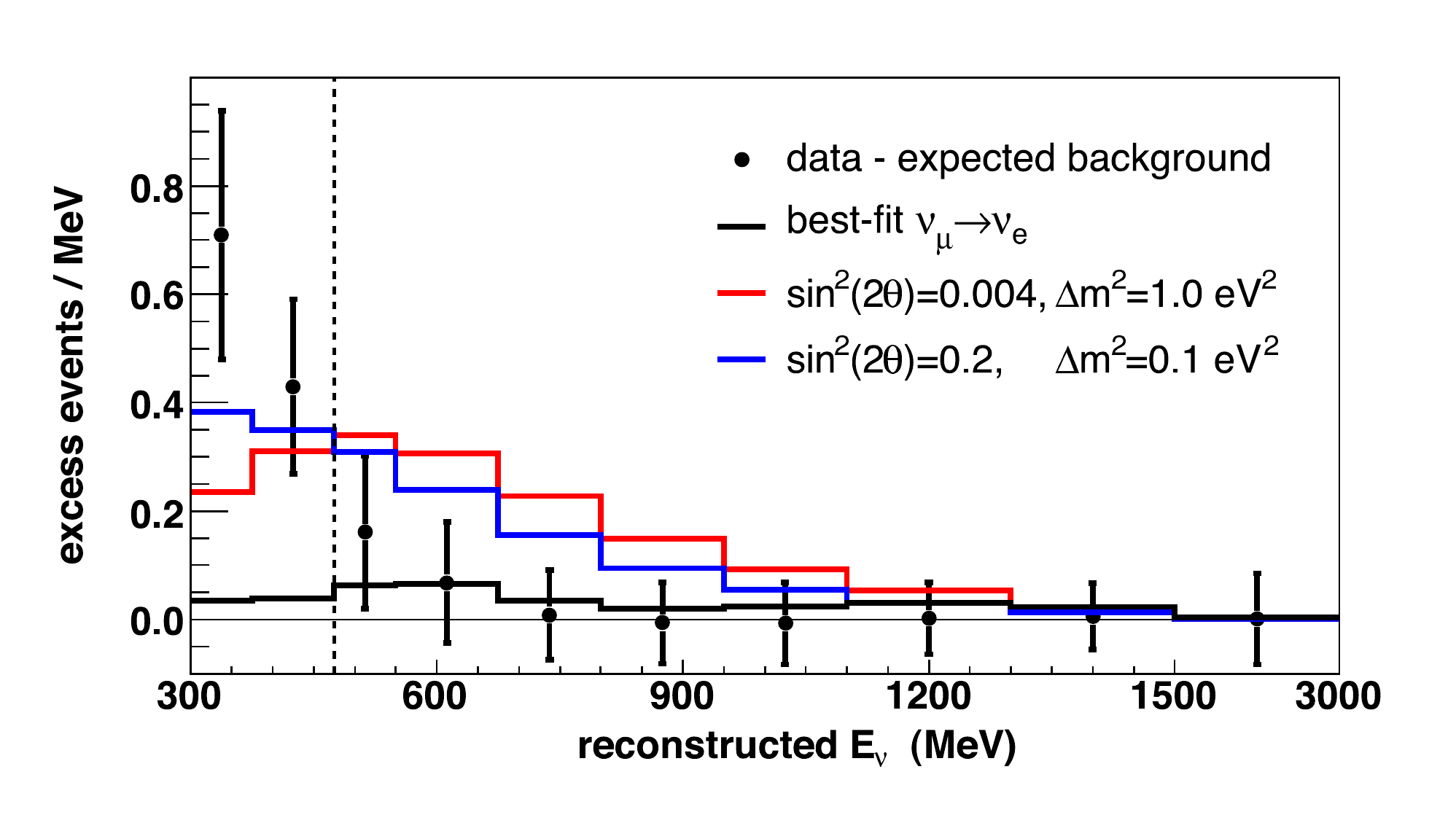}
\caption{Left: Observed $\enuqe$ distribution for selected $\nue$ CCQE candidate
events. The expected backgrounds are shown as stacked components. The
error bars on the predicted background represent the square root of the diagonal
components of the covariance matrix of the distribution. The dashed purple line shows
the expected $\enuqe$ distribution including the contribution of oscillation events from the best fit oscillation parameters. Right: The $\enuqe$ distribution of the  excess of $\nue$ CCQE candidate events, obtained by subtracting the predicted background from the observed distribution. The black line shows the $\enuqe$ distribution
of the best fit oscillation parameters, while the red and blue show the predicted excess
from two other oscillation parameters from the LSND allowed region. } \label{fig:results}
\end{figure*}

\section{Signal Extraction and Unblinding}
\label{sec:unblinding}

The $\nue$ signal is extracted by a $\chi^2$ fit to the $\enuqe$ distribution of the selected $\nue$ candidates. The observed energy distribution between 475
and $3000\mev$ is summarized in a histogram with eight bins, and
the  equivalent distribution for the expected background and signal for a given oscillation parameter set
($\sin^2 2\theta$, $\Delta m^2$) from the Monte Carlo simulation is created. The uncertainties in the background prediction and the signal are summarized in a covariance matrix. The best-fit $\sin^2 2\theta)$ and $\Delta m^2$ values
are extracted by varying these parameters and minimizing the $\chi^2$ between the observed data and the predicted background plus signal contributions. The signal extraction assumes 
that the energy distribution of the signal takes the form of neutrino oscillations driven by
one $\Delta m^2$.

The results of the analysis were unblinded in a staged procedure. The primary means of
checking the behavior of variables and distributions within the blinded events were a 
series of $\chi^2$ tests on geometric, kinematic and background suppression
variables using the total covariance from both statistical and systematic errors. 
Since the $\chi^2$ tests must allow for the presence of an excess of $\nue$ events from oscillation, the signal extraction fit is performed and the expected contribution from the fitted signal propagated into the distributions in addition to the background. The covariance matrix
for the distribution used in the $\chi^2$ tests are also updated. This procedure is performed without
reporting the results of the signal extraction fit. 

In the first step, only the $\chi^2$ from the  comparisons (with the exception of $\enuqe$) are reported without showing the data distributions, nor the expectations from the best fit. In the second step, the data and best-fit distributions in the variables (with the exception of $\enuqe$) are shown with the normalization and 
signal/background composition suppressed. In the final stage, the event yield, along with the
$\enuqe$ distribution and the best fit parameters are unblinded, completing
the unblinding the process. While the $\enuqe$ distribution is used to obtain the best fit
parameters, neither the $\chi^2$ probability of this fit nor the distribution are examined until
the final stage.

In the process of carrying out the first step using an energy range of $300<\enuqe<3000\mev$, it was found that the  $\chi^2$ for the visible energy distribution had a small probability of $0.01$. No other distributions were found to be problematic. Since the test accounts for the possibility of
a signal, the discrepancy could not be due to an excess of events consistent with neutrino
oscillations with a single $\Delta m^2$. A possible discrepancy was also observed in the
$\logepi/\mgg$ sideband samples described in Section \ref{sec:xcheck} in the form of an
excess of events over the predicted background at low energy. This discrepancy, however,
was not large enough to result in a poor $\chi^2$ for the energy distributions in these
control samples.  An examination of the unsigned deviation in the visible energy distribution of the observed data in the signal sample to the best-fit distribution confirmed that the poor $\chi^2$ was likely due to a discrepancy at low energy.
\begin{figure*}[t]

\centering
\includegraphics[width= 160.0 mm]{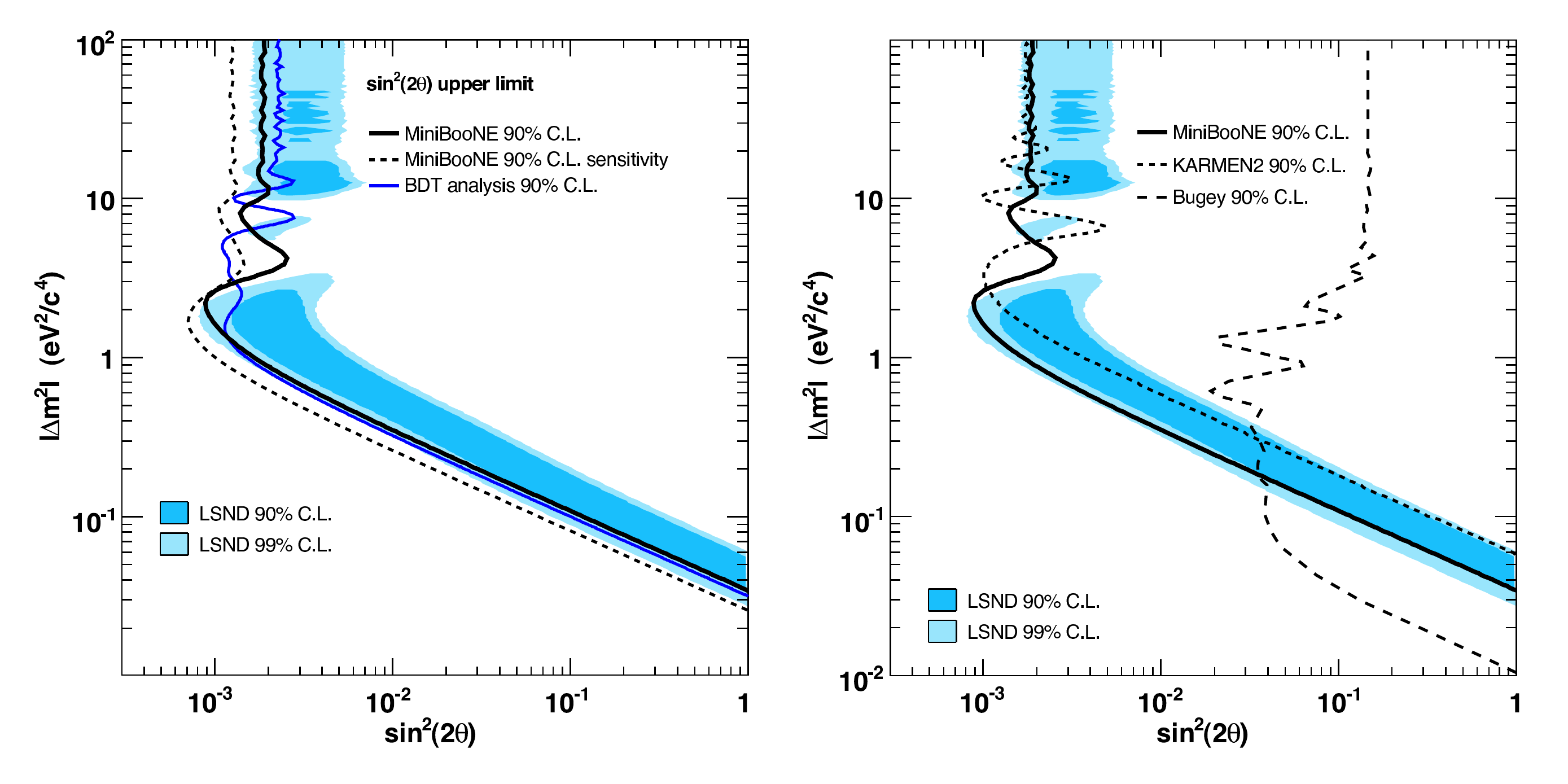}
\caption{Left: Limits at $90\%$ confidence level on the $\num\to\nue$ oscillation
parameters obtained from the MiniBooNE data using the one-dimensional 
raster scan method. The solid black line shows the
limit, while the dashed black line shows the projected sensitivity. The solid blue line
represents the limit obtained from the BDT analysis of the same data. Right: Limits
on $\num\to\nue$ oscillations  from MiniBooNE, KARMEN2 and Bugey. The
cyan and light blue regions on each plot are the parameter space compatible
at $90\%$ and $99\%$ confidence level with the LSND oscillation evidence, respectively. } 
\label{fig:limits}
\end{figure*}

Based on this information, the impact of increasing the $\enuqe$ threshold was investigated. It was
found that restricting the analysis to  $475<\enuqe<3000\mev$ did not impact
the sensitivity of the analysis to neutrino oscillations with a single $\Delta m^2$. After deciding
to implement this increased threshold in the analysis, the unblinding procedure was repeated.
No problems were found in this second iteration, with all distributions, including the visible
energy distribution, returning reasonable $\chi^2$ probabilities. As a result, the procedure
was taken to completion.

\section{Results}
The $\enuqe$ distribution of the signal candidates is shown on the left in Figure \ref{fig:results}.
In the analysis region of $475<\enuqe<3000\mev$, 380 events are observed, where the
expected background is $355\pm19(\mbox{stat})\pm35(\mbox{sys})$, corresponding
to an excess of $0.55$ standard deviations over background. The fit  yields a value of $(\sin^2 2\theta),\Delta m^2)= (1.1\times10^{-3},4.1\evsq)$. The background and signal
distribution corresponding to these parameters is shown as a dashed purple line. The difference
in $\chi^2$ between the null hypothesis and the best fit is 0.94, while the corresponding
 $\chi^2$ difference for the LSND best fit parameters is 13.7. This indicates that
 the former is an adequate fit to the data, while the latter is highly
 disfavored. In summary, both the overall yield of events and the $\enuqe$ distribution are consistent with the expected background.

A significant excess of events is observed at $300<\enuqe<475\mev$, below the analysis threshold,
as expected from the poor $\chi^2$s in the initial consistency tests described in
Section \ref{sec:unblinding}. The right plot in Figure \ref{fig:results} shows the excess of
data events over background. The black histogram shows the expected excess for the best
fit parameters, while the red and blue histogram show the excess expected for two different
oscillation parameters consistent with the LSND evidence, $(\sin^2 2\theta), \Delta m^2) = (0.004, 1.0 \evsq)$  and $(0.2, 0.1\evsq)$, respectively. While lower values of $\Delta m^2$ produce oscillation signatures which concentrate
at lower energies, it is not possible to accommodate the observed excess with neutrino oscillations
driven by a single $\Delta m^2$. This also follows from the fact that the fit did not find an adequate
solution when the low energy region was included. The excess at low energy is currently under investigation. 

The $90\%$ confidence level limits on the $\num\to\nue$ oscillation parameters are shown on the left in Figure \ref{fig:limits}, where parameters to the right of the solid black line are excluded.
 The limits are obtained in a one-dimensional raster scan whereby a limit
on the maximum allowed $\sin^2 2\theta)$ is determined at each $\Delta m^2$ to obtain the curve.
For comparison, the dashed line indicates the projected sensitivity of the analysis. The obtained limit is somewhat worse
than the projected sensitivity due to the small excess of events observed in the data.
The limits from the BDT analysis, which was unblinded simultaneously and did not
observe a significant excess of events, are also shown in Figure \ref{fig:limits}. In addition to
a different event selection scheme, the BDT analysis
also used a different signal extraction procedure in which the $\num$ and $\nue$ CCQE samples were fit simultaneously.

\section{Conclusions}
MiniBooNE has searched for $\num\to\nue$ oscillations in a sample
of  $5.58\times 10^{20}$ protons-on-target delievered to the Booster Neutrino Beam
in  neutrino mode.
The primary backgrounds to the analysis are constrained with {\em in situ}
measurements and cross checks. The analysis yields no evidence for
neutrino oscillations: the observed yield and energy spectrum of 
the selected $\nue$ candidates are consistent with the background and incompatible
with the oscillations indicated by the LSND experiment. An excess of events
at energies below the analysis $\enuqe$ threshold of $475\mev$ remains
under investigation. The oscillation parameters excluded by the analysis
are shown on the right in Figure \ref{fig:limits}, together with the allowed
regions from the LSND analysis and limits from the KARMEN and Bugey
experiments.

\begin{acknowledgments}
The MiniBooNE collaboration acknowledges support from the Department of
Energy and the National Science Foundation of the United States. We are
grateful to Fermilab for hosting the experiment and for the excellent 
accelerator performance. We thank  Los Alamos National Laboratory for LDRD funding.
We acknowledge Bartoszek Engineering for the design of 
the focusing horn. We acknowledge Dmitri Toptygin, Anna 
Pla, and Hans-Otto Meyer for optical measurements of 
mineral oil. This research was done using resources provided by the Open Science Grid, which is supported by the 
NSF and DOE-SC. We also acknowledge the use of the 
LANL PINK cluster and CONDOR software in the analysis of 
the data. 

\end{acknowledgments}

\bigskip 

\end{document}